\documentclass [aps,prb,twocolumn,amssymb,amsmath,showpacs,floatfix,superscriptaddress]{revtex4-2}

\makeatletter
  \long\def\pprintMaketitle{\clearpage
  \iflongmktitle\if@twocolumn\let\columnwidth=\textwidth\fi\fi
  \resetTitleCounters
  \def\baselinestretch{1}%
  \printFirstPageNotes
  \begin{center}%
 \thispagestyle{pprintTitle}%
   \def\baselinestretch{1}%
    \Large\@title\par\vskip18pt
    \normalsize\elsauthors\par\vskip10pt
    \footnotesize\itshape\elsaddress\par\vskip36pt
    \end{center}%
  \gdef\thefootnote{\arabic{footnote}}%
  }
\makeatother

\usepackage{amsmath}
\usepackage{amssymb}
\usepackage{amsfonts}
\usepackage{graphicx}
\usepackage{verbatim}
\usepackage{bm}
\usepackage{mathbbol}
\usepackage{color}
\setcitestyle{compress}

\usepackage{mathtools}

\usepackage{tabularx}
 
\usepackage[normalem]{ulem}

\usepackage{graphicx,epsfig,amsfonts,amssymb}
\usepackage{bm}
\usepackage{times}
\usepackage{lipsum}
\usepackage{verbatim}

\usepackage{hyperref}
\hypersetup{
    colorlinks,
    citecolor=blue,
    filecolor=blue,
	    linkcolor=blue,
    urlcolor=blue
}

\newcommand{\beg}{\begin{equation}}
\newcommand{\en}{\end{equation}}

 \newcommand{\lam}{\lambda}

\newcommand{\eref}[1]{Eq.~(\ref{#1})}
\newcommand{\re}[1]{(\ref{#1})}

\newcommand{\esref}[1]{Eqs.~(\ref{#1})}
\renewcommand{\Re}{\mathrm{Re}}
\renewcommand{\Im}{\mathrm{Im}}

\newcommand{\eps}{\varepsilon}

\newcommand{\dn}{\downarrow}
\newcommand{\up}{\uparrow}

\newcommand{\pp}{{\bf p}}

\newcommand{\ps}{\mathsf{p}}

\begin{document}

\title{Superconductivity near a quantum critical point in the extreme retardation regime}

\author{Emil A. Yuzbashyan}
\address{Department of Physics and Astronomy, Rutgers University, Piscataway, New Jersey 08854, USA}

\author{Michael K.-H. Kiessling}
\address{Department of Mathematics, Rutgers  University, Piscataway, New Jersey 08854, USA}

\author{Boris L. Altshuler}
\address{Physics Department, Columbia University, 538 West 120th Street, New York, New York 10027, USA}

\begin{abstract}

We study     fermions at quantum criticality    with extremely retarded interactions of the form $V(\omega_l)=(g/|\omega_l|)^\gamma$, 
where $\omega_l$ is the transferred Matsubara frequency. This system undergoes a normal-superconductor phase transition at a critical temperature $T=T_c$. The order parameter is the frequency-dependent gap function $\Delta(\omega_n)$ as in the Eliashberg theory. In general,  the interaction is extremely retarded for $\gamma\gg 1$, except at low temperatures    $\gamma>3$ is sufficient. We evaluate the normal state specific heat, $T_c$, the jump in the specific heat, $\Delta(\omega_n)$ near $T_c$, and the Landau free energy. Our answers  are asymptotically exact in 
the limit $\gamma\to\infty$. At low temperatures, we prove that the global minimum of the free energy is nondegenerate and determine the order parameter,  the free energy, and the specific heat. These answers are exact for $T\to0$ and $\gamma>3$. We also uncover and investigate an instability of the $\gamma$ model: negative specific heat at $T\to0$ and just above $T_c$.  

\end{abstract}

\maketitle

\section{Introduction}

Critical fluctuations of the order parameter field mediate strong electron-electron interactions near metallic quantum critical points. 
Coupling to this bosonic order parameter field induces attraction between electrons, leading to the enhancement of superconductivity near  such points in many quantum materials, such as the cuprates, iron pnictides, and certain heavy-fermion materials~\cite{broun,matsuda,stewart,sri}.
On the other hand, same  interactions  destroy  Fermi liquid quasiparticles thus undermining superconductivity which builds on   quasiparticle coherence. As a result, the enhancement   does not occur in other quantum critical materials, such as
heavy-fermion metals CeCu$_{6-x}$Au$_x$ and YbRh$_2$Si$_2$~\cite{vojta}.

It has been argued~\cite{andrey1}  that many quantum critical system are described by  the \textit{$\gamma$ model} -- effective electron-electron interactions   of the form $V(\omega_l)=(g/|\omega_l|)^\gamma$, where $\omega_l=2\pi T l$ is the transferred Matsubara frequency. Examples include   a 2D nematic ($\gamma=1/3$)  and magnetic ($\gamma=1/2$) quantum critical points, a spin-liquid model for the cuprates ($\gamma=0.7$), 2D pairing mediated by an undamped propagating boson ($\gamma=1$), and  the strong coupling limit of phonon mediated superconductivity ($\gamma=2$). Interactions of this form obtain by integrating out  the bosonic fields and  
averaging   over the Fermi surface, 
taking into account that the order parameter field   is massless at criticality.

Here we study  the thermodynamic properties of   the  $\gamma$ model   in the extreme retardation regime, when   $V(\omega_l)$ is effectively local in the Matsubara frequency domain. This corresponds to $\gamma>3$ at low temperatures  and to $\gamma\gg1$ at arbitrary $T$.
 Even though in all the above examples $0\le \gamma\le 2$, there is no fundamental reason why larger $\gamma$ are impossible. Moreover, properties of the system at finite temperature are continuous in $\gamma$. For example, the relative difference between the exact superconducting transition temperature $T_c$ for $\gamma=2$ and the large $\gamma$ asymptote for $T_c$ is 
only 6\% (see below). 

The attractive feature of the $\gamma$ model is that there is a single intrinsic dimensionless parameter in the model -- the exponent $\gamma$ itself. The coupling constant $g$ sets the energy units. The momentum-dependence of the fermion-boson coupling and the spectrum of mediating bosons become irrelevant at criticality. The Fermi energy is   the largest energy in the problem, i.e., we are in the limit $\eps_F/g\to\infty$. This suggests that the $\gamma$ model is the minimal description of the strong coupling fixed point (quantum critical point) in certain materials.

The interaction as we wrote it  diverges at $\omega_l=0$. Moving slightly away from criticality regularizes this divergence. The particular form of the regularization is unimportant, because it does not affect our  results. A convenient regularization is $V(\omega_l)=\frac{g^\gamma}{|\omega_l|^\gamma+\Omega^\gamma}$, where $\Omega$ has the meaning of the critical boson mass. Now it is straightforward to analyze the interaction in the time domain. In the $\gamma\to\infty$ limit, $V(0)= \frac{g^\gamma}{\Omega^\gamma}\equiv\lam$  and $V(\omega_l)=0$ for $\omega_l\ne0$. Its Fourier transform is $V(\tau'-\tau)= \lam T$. This interaction is  \textit{maximally retarded} --  independent of the time separation between the interacting particles both in real and imaginary time domains. By contrast, instantaneous  interactions, e.g., the pairing interaction   in the Bardeen-Cooper-Schrieffer (BCS) theory, are proportional to $\delta(\tau'-\tau)$. When $\gamma$ is large but finite, the interaction is \textit{extremely retarded}.

The $\gamma$ model is a generalization of the strong coupling limit of the Eliashberg theory~\cite{carbotte,combescot,spinchain}, which corresponds to $\gamma=2$. It presents two phases: normal and superconducting. The order parameter is the frequency dependent gap function $\Delta_n\equiv\Delta(\omega_n)$ as in the Eliashberg theory. Here $\omega_n=\pi T(2n+1)$ is the fermionic Matsubara frequency. We focus on the thermodynamic properties near the superconducting transition temperature  $T_c$ and
at low $T$. Our results near $T_c$ are asymptotically exact in the limit $\gamma\to\infty$ and the low temperature answers are similarly exact for $T\to0$ at any $\gamma>3$.   

First, we  determine the $T_c$ itself, 
\beg
T_c(\gamma)=\frac{ g a^{\frac{1}{\gamma}}}{2\pi},\quad a\approx 1.1843,
\label{tc_intro}
\en
 and then the order parameter near $T_c$
 \beg
 \Delta_n=\eta\psi a^{-\frac{1}{\gamma}}|\omega_n| J_{n+\frac{1}{2}+a^{-1}}(a^{-1}),
 \en
 where $\eta\approx19.20$, $J_\alpha(x)$ is the Bessel function of the first kind, and  the complex number  $\psi$ is the Landau order parameter.  Note that   \eref{tc_intro}  appeared  in an earlier paper~\cite{2mats}.
 
 The free energy expanded   to the fourth power of $|\psi|$ near the transition (the Landau free energy) is
 \beg
f_L = R \nu_0 g^2 \left [\frac{2\pi\gamma}{g} (T-T_c) |\psi|^2+\frac{|\psi|^4}{2}\right],
\label{conden_intro}
\en
where $R\approx 0.62$ and  $\nu_0$ is the density of states at the Fermi level. This allows us to evaluate  $|\psi|$ near $T_c$,
\beg
|\psi|=  \left[\frac{2\pi\gamma}{g}\right]^{1/2} (T_c-T)^{1/2},
\en
the jump in the specific heat at $T_c$,
\beg
\delta c=c_\mathrm{s}-c_\mathrm{n}=2\pi R\nu_0 g \gamma^2 a^{\frac{1}{\gamma}},
\label{Cjump_intro}
\en
and the thermodynamic critical field,
\beg
H_c= 4\pi^{3/2} \gamma \sqrt{R\nu_0} (T_c-T).
\en
As expected $|\psi|$ and $H_c$ have the usual mean-field critical exponents, but note also their $\gamma$-dependence.   

  We prove that  the global minimum of the free energy is unique for $T\to0$ up to an overall phase $e^{i\phi}$ of the 
gap function $\Delta(\omega_n)$ and determine the leading small $T$ asymptotic behavior of  $\Delta(\omega_n)$. Specifically,
\beg 
 \frac{ \Delta(\omega_n)}{\omega_*}=Y\!\left(\frac{\omega_n}{\omega_*}\right), \quad Y(x)= x \tan \theta_0(x),
 \label{delta_intro}
\en
where
\beg
 \omega_* =g \frac{[\zeta(\gamma-2)]^{\frac{1}{3}}}{2^{\frac{1}{3}} } \left(\frac{g}{2\pi T}\right)^{\!\!\frac{\gamma}{3}-1}\!\!\!\!\!\!\!\!\!\!,
 \en
and $\theta_0(x)$ is the solution of the universal gap equation $\theta''= x \sin\theta$. Equation~\re{delta_intro} says that
plots of the gap functions $\Delta(\omega_n)$  vs   $\omega_n$  collapse onto the same curve (Fig.~\ref{deltafig} below) for all $\gamma>3$, $g$, and  $T\to0$, when both $\Delta(\omega_n)$ and $\omega_n$ are measured in units of $\omega_*$. Further, we  evaluate $2\Delta(0)/T_c$ and find that it is finite  for all $\gamma>3$ and  $T>0$, but diverges as $|\gamma-3|^{-1/3}$ for  $\gamma\to 3^+$ and as $T^{1-\gamma/3}$ for $T\to0$.

We also determine the normal state specific heat $c_\mathrm{n}$ at all temperatures and the  specific heat $c_\mathrm{s}$ in the superconducting state just below $T_c$ and for $T\to0$. The analysis  of the specific heat reveals a pathology in the $\gamma$ model: $c_\mathrm{n}$ is negative just above $T_c$ and $c_\mathrm{s}$ is negative at $T\to0$. This indicates that the $\gamma$ model is thermodynamically unstable~\cite{landau} at these temperatures. The resolution  of this problem depends on the microscopic Hamiltonian
underlying the $\gamma$ model, which is not given. We discuss several scenarios. In particular, there  is a scenario that removes this pathology without affecting any of the above answers.

The content of this paper is as follows. In Sec.~\ref{mapping_sec}, we derive the free energy functional   of the $\gamma$ model and
map it to a classical Heisenberg spin chain  using the approach we proposed in an earlier paper. In Sec.~\ref{stpt}, we obtain the equation for the stationary points of the effective action for arbitrary $\gamma$ and in Sec.~\ref{extreme_sec} we specialize   the free energy  to the extreme retardation regime. In Sec.~\ref{sctran_sec}, we first device a numerical approach for determining $T_c$ for arbitrary $\gamma$ and then obtain   the solution of the linearized gap equation for 
$\gamma\to\infty$. We derive the large $\gamma$ asymptote of $T_c$ in Sec.~\ref{crit_T_sec}. Section~\ref{thermo_sec} contains our results for the Landau free energy,  jump
in the specific heat at $T_c$,   Landau order parameter, and  thermodynamic critical field. We work out
the  specific heat in the normal state and in the superconducting state just below $T_c$ in Sec.~\ref{ns_sec}. In Sec.~\ref{lowTsec},
we solve the gap equation at low temperatures, and evaluate the specific heat in the superconducting state at $T\to0$. In the final section, we  summarize and pose an important open problem. The Appendix provides a   proof of the   uniqueness of  the global minimum of the free energy at small $T$.

\section{Mapping to the spin chain}
\label{mapping_sec}

In previous paper~\cite{spinchain} we mapped the effective action for electrons interacting via phonons to a classical spin chain. In the strong coupling limit where phonon frequencies go to zero, this is the $\gamma$ model with $\gamma=2$.  The same approach works for arbitrary $\gamma$. We define the $\gamma$ model by its Euclidean  action
\beg
\begin{split}
S=T\sum_{ \ps\sigma} & \psi^*_{\ps\sigma} G_{\ps}^{-1} \psi_{\ps\sigma}-\\
  &\frac{T^3\delta}{2}\!\!  \sum_{ \ps_i\sigma\sigma'} 
 \frac{  g^\gamma }{|\omega_l|^\gamma+\Omega^\gamma}  \psi^*_{ \ps_1\sigma} \psi_{ \ps_3\sigma}  \psi^*_{ \ps_2\sigma'} \psi_{ \ps_4\sigma'},
\end{split}
\label{postulate}
\en
where $G_{\ps}^{-1}= -i\omega_n+\xi_\pp$,  $\omega_n$ are fermionic Matsubara frequencies, $\xi_\pp$ are  single-particle levels counted from the chemical potential, $\psi_{\ps\sigma}$ is the fermionic field,  and $\ps_i$ are  frequency-momentum 4-vectors constrained
by $\ps_1-\ps_3=\ps_2-\ps_4=(\omega_l, \bm q)$.  The single-particle level spacing $\delta$ is a combination of the density of states at the Fermi energy $\nu_0$ and  the system volume $V$, $\delta=(\nu_0 V)^{-1}$. We added a constant  $\Omega\ll T$ to avoid divergence at zero frequency.  In the end we take the limit $\Omega\to 0$.

This action is exact for the electron-phonon system ($\gamma=2$) in the limit  when phonon frequencies go to zero, the Fermi energy $\eps_F\to\infty$, and assuming the electron-phonon coupling depends only on the magnitude of the transferred momentum. Its status is unclear in most other examples listed above, when the bosonic field is a collective mode containing fermions themselves, such as, e.g., the nematic or magnetic order parameter. In this paper, we simply postulate~\eref{postulate} and study its properties regardless of its origin.

The mapping to the spin chain involves several steps~\cite{spinchain}. First, we decouple the interaction with three Hubbard-Stratonovich fields $\Sigma_\up$, $\Sigma_\dn$, and $\Phi$ and integrate out the fermions~\cite{couple}. Stationary point equations for these fields are    the Eliashberg equations generalized to arbitrary $\gamma$ from $\gamma=2$.  Fluctuations around the stationary point are negligible as long as the Fermi energy is the largest energy in the problem~\cite{meaningmigdal}.  To unveil the spin chain, we expand the action   in $\Omega$ and then send $\Omega\to0$~\cite{alternative}.  A key  ingredient of the mapping is  the observation  that at the stationary points the normal   and anomalous  Green's functions integrated over $\xi_\pp$ satisfy the constraint $G_n^2+|F_n|^2=1$. The normal average $G_n$ is real and the anomalous average $F_n$ is complex. Moreover, the mass of fluctuations violating this constraint is proportional to $\Omega^{-\gamma}\to\infty$~\cite{meaningmigdal}. The constraint is therefore rigidly enforced even  away from the stationary point. 

\begin{figure}
	\centering
	 \includegraphics[width=0.9\columnwidth]{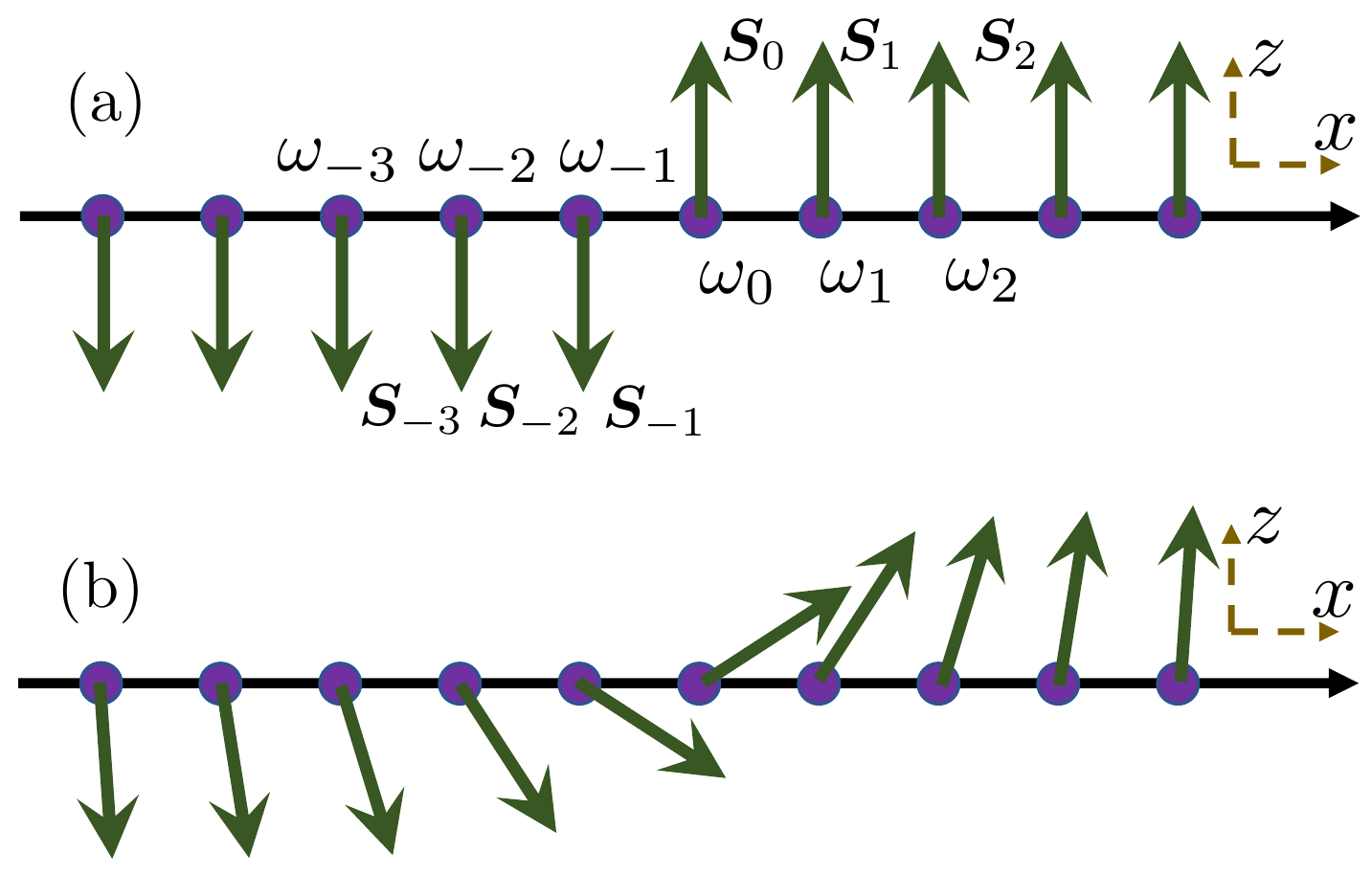}
	\caption{Transition from (a) the normal state to (b) a superconducting state in terms of classical spins.  Positions of the spins $\bm S_n$ are the fermionic Matsubara frequencies $\omega_n$. Spin-spin interactions are purely ferromagnetic and the spins are subject to a Zeeman magnetic field  $2\pi\omega_n$  along the $z$ axis.     In the superconducting state, spins acquire  $x$-components, which implies nonzero anomalous  Green's function. The normal-superconductor transition is seen as the softening of the sharp domain wall at the origin of the spin chain.  }
	\label{spinsfig}
\end{figure}

It is now natural to introduce  classical spins $\bm S_n$ of unit length as
\beg
S_n^z=G_n,\quad S_n^x=\Re (F_n),\quad S_n^y=\Im (F_n).
\label{sgf}
\en
In terms of  the spins, the action expanded in $\Omega$ becomes $S_\mathrm{eff}=\nu_0 V H_s$, where $H_s$ is the spin chain Hamiltonian  
\beg
H_s=-2\pi \sum_n \omega_n S_n^z-\pi^2 Tg^\gamma\sum_{nm} \frac{ {\bm S_n}\cdot {\bm S_m}-1}{|\omega_n-\omega_m|^\gamma}.
\label{sch}
\en
In addition to the expansion in $\Omega$, we regularized the interaction term~\cite{spinchain}  by replacing ${\bm S_n}\cdot {\bm S_m}\to{\bm S_n}\cdot {\bm S_m}-1$ and
then took the limit $\Omega\to 0$ which is now harmless. The free energy density of a given field configuration is~\cite{grand}
\beg
f = \frac{ T S_\mathrm{eff}}{V}=\nu_0 T H_s.
\label{freespin}
\en
The Boltzmann weight of a given field configuration   is $e^{-f V/T} =e^{-H_s/\delta}$, i.e., the spin chain is at an effective temperature
$T_s=\delta$. We work in the thermodynamic limit where $\delta\to 0$ and therefore the spin chain is at zero temperature.

  We introduce  the frequency-dependent superconducting gap function $\Delta_n\equiv\Delta(\omega_n)$ through~\cite{spinchain,spinflip}
\beg
S_n^z=\frac{ \omega_n}{\sqrt{\omega_n^2+|\Delta_n|^2}},\quad S_n^+=\frac{ \Delta_n}{\sqrt{\omega_n^2+|\Delta_n|^2}},
\label{spintogap}
\en
where $S_n^+=S_n^x+i S_n^y$. 

In the normal state, $\Delta_n=0$ and as a consequence  $\bm S_n=\mbox{sgn} (\omega_n) \hat z$, where $\hat z$ is the unit vector along the $z$ axis. A characteristic feature of this state is a sharp domain wall between $\omega_{-1}$ and $\omega_0$. In the superconducting state,   $\Delta_n\ne0$ and the domain wall softens, see Fig.~\ref{spinsfig}.

\section{Stationary points of the free energy}
\label{stpt}

In Ref.~\cite{spinchain}, we discussed  stationary points of the free energy functional $f$ for $\gamma=2$. These results in fact apply to any $\gamma>1$.  In particular, we know  that with a proper choice of the arbitrary overall phase $e^{i\phi}$, the gap function at the minimum of the free energy has the following properties: $\Delta(\omega_n)> 0$ for all $\omega_n$ or $\Delta(\omega_n)=0$ for all $\omega_n$,  $\Delta(\omega_n)$ is an even function,  and $\Delta(\omega_n)\to0$ as $\omega_n\to\pm\infty$. These properties hold for  $T=0$ as well, in which case $\omega_n$ becomes a continuous variable $\omega$ and $\Delta(\omega_n)> 0$ implies  $\Delta(0)>0$. 
 We also know that  the stationary points of $f$ and of the spin chain $H_s$ are identical, because we are in the strong coupling limit  $\lam=\frac{g^\gamma}{\Omega^\gamma}\to\infty$ (equivalent to $\Omega\to0$). 
 
 Given that $\Delta_n\ge 0$, \eref{sgf} implies $S_n^y=0$ and $S_n^x\ge 0$. Let $\theta_n\equiv\theta(\omega_n)$ be the angle $\bm S_n$ makes with the  $z$ axis. Then
\beg
S_n^z=\cos\theta_n,\quad S_n^x=\sin\theta_n,\quad  0\le\theta_n\le \pi.
\label{spintheta}
\en
The angle $\theta_n$ cannot exceed $\pi$ due to the condition $S_n^x\ge0$. This also prevents $\theta_n$ from winding.
The remaining properties of the gap function
\beg
\Delta_n=\omega_n\tan\theta_n
\label{deltatheta}
\en
 translate into
\beg
\theta(-\omega_n)=\pi-\theta(\omega_n),\quad \lim_{\omega_n\to\pm\infty} \theta_n=\frac{\pi}{2}\mp \frac{\pi}{2}.
\label{sym}
\en

The spin chain Hamiltonian reads in terms of $\theta_n$
$$
H_s=-2\pi \sum_n \omega_n \cos\theta_n-\pi^2 Tg^\gamma\sum_{nm} \frac{ \cos(\theta_n-\theta_m)-1}{|\omega_n-\omega_m|^\gamma}.
 $$
 Differentiating this with respect to $\theta_n$, we obtain the equation for the stationary points of the free energy of the $\gamma$ model
 \beg
 \omega_n\sin\theta_n= g^\gamma \pi T \sum_{m\ne n} \frac{\sin(\theta_m-\theta_n)}{|\omega_m-\omega_n|^\gamma}.
 \label{eligapeq}
 \en
  This is the superconducting \textit{gap equation}. It is a version of the Eliashberg gap equation as it follows from the generalized Eliashberg equations mentioned in the previous section. It takes the form of the standard   gap equation for
  the strong coupling limit of the Eliashberg theory~\cite{combescot,carbotte} when we rewrite it in terms of $\Delta_n$ and set $\gamma=2$.
  
  The \textit{normal state}, $\Delta_n=0$,  is always a stationary point. The corresponding spin configuration  $\bm S_n=\mbox{sgn} (\omega_n) \hat z$  is shown in  Fig.~\ref{spinsfig}(a). In  this state  $\theta_n=0$ for $\omega_n>0$ and $\theta_n=\pi$ for $\omega_n<0$, which indeed satisfies the stationary point equation~\re{eligapeq}. It is also not difficult to show that since $S_n^z$ is the normal Green's function integrated over $\xi_\pp$,
 in the normal state all states below the chemical potential  are occupied and above -- empty. At $T=0$, the Matsubara frequency $\omega_n\to\omega$  takes values on the entire real axis.  The angle $\theta(\omega)$ is then discontinuous at $\omega=0$ in the normal state. In contrast, in the ground state (free energy minimum at $T=0$) we have
 \beg
 \theta(0)=\frac{\pi}{2}.
 \label{theta_gr_st}
 \en
 This follows from \esref{spintogap} and \re{spintheta} together with the condition $\Delta(0)>0$.
    
  The symmetry~\re{sym} enables us to express the free energy density~\re{freespin} in terms of $\theta_{n\ge0}$ only. 
It is also convenient to introduce along the way the non-dimensionalized free energy density $\bar f$  
	  as follows:
	  \begin{equation}
  \begin{split}
& \bar f\equiv \frac{f}{\nu_0 g^2}=-2\tau \sum_{n=0}^\infty  \bar\omega_n\cos\theta_n-\\
& \frac{\tau^{2-\gamma}}{2}\!\!\!\!\sum_{n,m=0}^\infty\!\left[
 \frac{\cos(\theta_n-\theta_m)-1}{(n-m)^\gamma}-
\frac{\cos(\theta_n+\theta_m)+1}{(n+m+1)^\gamma}\right]\!,
\end{split}
\label{freesym}
\end{equation}
where the dimensionless temperature $\tau$ and fermionic Matsubara frequency $\bar\omega_n$ are
\beg
\tau =\frac{2\pi T}{g},\quad \bar\omega_n=\frac{\omega_n}{g}=\tau\left(n+\frac{1}{2}\right)\!\!.
\label{omegabar}
\en
  
 \section{Extreme retardation regime}
 \label{extreme_sec}
 
 In this paper we are primarily interested in the extreme retardation regime of the $\gamma$ model. In this regime, the range of the interaction is extremely short  in the frequency domain and therefore extremely long (retarded) in the time domain. In terms of the spin chain~\re{sch}, this means that a given spin $\bm S_n$ interacts only with its nearest neighbors or, in the continuum $T\to0$ limit, only
 with spins in its immediate vicinity.  
  There are two overlapping cases when this happens. One is
 $\gamma\gg1$ at arbitrary $T$. The other is low $T$ for any $\gamma>3$. Let us consider these two cases.
 
 \subsection{Large $\gamma$, arbitrary $T$}
 
 At large $\gamma$, the $|n-m|\ge 2$ interaction terms  in \eref{sch} are suppressed  by a factor of $2^{-\gamma}$ or smaller compared to the $|n-m|=1$ terms.
Discarding exponentially small terms, we obtain a spin chain with nearest neighbor interactions, 
\beg
H_s=-2\pi \sum_n \omega_n S_n^z-  \frac{\pi  g^\gamma}{(2\pi T)^{\gamma-1}}\sum_{n}  ( {\bm S_n}\cdot {\bm S_{n+1}}-1).
\label{sch2}
\en
The corresponding  non-dimensionalized free energy density  for the coplanar spin distribution~\re{spintheta} is
\beg
\begin{split}
\bar f\equiv \frac{f}{\nu_0 g^2}= &-\tau \sum_n\bar \omega_n \cos\theta_n- \\
&\frac{\tau^{2-\gamma} }{2}\sum_{n} \left[   \cos(\theta_{n+1}-\theta_n)-1\right],
\end{split}
\label{nonf}
\en
 and the expression~\re{freesym} in terms of  $\theta_{n\ge0}$ becomes
\beg
\begin{split}
\bar f=-2\tau \sum_{n\ge0} & \bar \omega_n \cos\theta_n +\tau^{2-\gamma}\frac{\cos(2\theta_0)+1}{2}\\
&-  \tau^{2-\gamma}  \sum_{n\ge0} \left[   \cos(\theta_{n+1}-\theta_n)-1\right].
\end{split}
\label{nonfpos}
\en
We use  this expression for the free energy  in subsequent sections to  determine the critical temperature $T_c$, the jump in the specific heat at the transition, etc.

\subsection{Any $\gamma>3$, low $T$}
\label{anyg3}

We will see below that at $T\ll g$ the $\gamma$ model is local for any $\gamma>3$. 
The gap equation becomes a nonlinear ordinary differential equation (ODE), and
the interaction part of the free energy corresponds to the continuum limit of the classical ferromagnetic 
Heisenberg spin chain with nearest-neighbor interactions. Again,  the interaction is 
as local in the frequency domain as it can be.

 Consider the solution $\theta_n=\theta(\omega_n)$ of  the gap equation~\re{eligapeq} under the conditions \re{sym}. One can always extend $\theta(\omega_n)$ to an infinitely differentiable function $\theta(\omega)$ for all real $\omega$
that coincides with $\theta_n$ when $\omega=\omega_n$, and such that conditions \re{sym} hold with $\omega$ in place of $\omega_n$.
  This also extends $\Delta(\omega_n)$ to $\Delta(\omega)$ with the help of \eref{deltatheta} and fixes 
  $\theta(0)=\frac{\pi}{2}$~\cite{interesting} by continuity of $\theta(\omega)$. There are uncountably many such  functions but we will determine a distinguished $\theta(\omega)$
that is   real analytic on the entire real $\omega$ axis~\cite{confused} and captures the small $T$ asymptotic behavior of $\theta(\omega_n)$.      We will see in Sec.~\ref{lowTsec} that $\theta(\omega)$ varies
 on a scale $\omega_*\propto T^{1-\frac{\gamma}{3}}$. The difference $|\omega_m-\omega_n|$ in \eref{eligapeq} is of the order of $\omega_*$
 when $|m-n|\sim (T/g)^{-\frac{\gamma}{3}}\gg1$.   Terms with  $|m-n|\gg1$ are negligible  due to  the rapid convergence of the sum over $m$. Therefore, $\theta(\omega_m)$  is close to $\theta(\omega_n)$ for $m$ whose contribution 
  to \eref{eligapeq} is  significant  and  
\beg
\begin{split}
\sin(\theta_m-\theta_n)\approx \theta_m-\theta_n \phantom{+\frac{d\theta_n}{d\omega_n}(\omega_m-\omega_n)+} \\ 
\approx
 \frac{d\theta_n}{d\omega_n}(\omega_m-\omega_n)+\frac{1}{2}\frac{d^2\theta_n}{d\omega_n^2}(\omega_m-\omega_n)^2.
 \end{split}
 \label{repl}
\en
Substituting \eref{repl} into the gap equation~\re{eligapeq} and dropping the discrete index $n$ [$\omega_n\to\omega$ and $\theta_n\to\theta(\omega)$], we find the nonlinear ODE
\beg
\bar\omega\sin\theta= \frac{\tau^{3-\gamma}}{2}\frac{d^2\theta}{d\bar\omega^2} \sum_{k=1}^\infty \frac{1}{k^{\gamma-2}},
\label{ode}
\en
where $\bar\omega=\omega/g$. We see that the approximations made require $\gamma>3$ and low $T$ 
so that the summation over $k$ converges and the replacement~\re{repl} is accurate. 
More precisely, in Sec.~\ref{lowTsec}
we show that the local  approximation~\re{ode} for the gap equation  and the corresponding approximation for the free energy  are accurate when $\tau^{\frac{r\gamma}{3}}\ll1$, where $r=\gamma-3$ for
$3<\gamma\le 5$ and $r=2$ for $\gamma\ge 5$. We use  \eref{ode} in Sec.~\ref{lowTsec}  to  determine the scaling laws at low $T$.

The  non-dimensionalized  free energy density in this approximation becomes
\beg
\begin{split}
\bar f = -\tau \sum_n\bar \omega_n \cos\theta_n+ 
\zeta(\gamma-2)\frac{\tau^{4-\gamma} }{4}\sum_{n}  \left(\frac{d\theta_n}{d\bar\omega}\right)^{\!\!2}\!\!,
\end{split}
\label{nonf22}
\en
where $\zeta(x)$ is the Riemann zeta function. The variation of this expression with respect to $\theta_n$ gives \eref{ode}. It is also
instructive to compare formulas \re{nonf22} and  \re{nonf} for the free energy. The low temperature limit of 
\eref{nonf} and the large $\gamma$ limit of \eref{nonf22} should coincide. This is indeed the case, because $1-\cos(\theta_{n+1}-\theta_n)\approx \frac{1}{2}\tau^2 (d\theta_n/d\bar\omega_n)^2$  at low $T$ and $\zeta(\gamma-2)\to 1$   as $\gamma\to\infty$. 

In terms of spins the free energy~\re{nonf22} takes a particularly simple form,
\beg
\bar f = \kappa^{-2} \int_{-\infty}^\infty dx\left\{ -x S^z +\frac{1}{2} \left( \frac{ d \bm S}{d x}\right)^{\!\!2} \right\}\!,
\label{freelow}
\en
where
\beg
\kappa^3=\frac{2\tau^{\gamma-3}}{\zeta(\gamma-2)},\quad x=\kappa\bar\omega.
\label{resc}
\en
This is the continuum limit of the classical ferromagnetic Heisenberg spin chain with nearest neighbor interactions and a Zeeman field. Note that the  free energy diverges as $T^{2-\frac{2\gamma}{3}}$ in the $T\to0$ limit. This divergence is cutoff by the mass $\Omega$ of the critical boson. In other words, the $\gamma$ model requires an infrared cutoff for $\gamma>3$. Physically,  we are working in the regime $T\gg\Omega>0$ and  by low $T$ we mean
$g\gg T\gg \Omega$.
  
 \section{Superconducting transition}
 \label{sctran_sec}
 
Here we first investigate the superconducting transition in the $\gamma$ model for \textit{arbitrary} $\gamma$ and then turn to $\gamma\to\infty$ limit.
 As mentioned in Sec.~\ref{stpt}, the normal state is a stationary point of the free energy functional at any temperature. It is the global minimum above a certain (critical) temperature $T_c$. Below $T_c$, which is nonzero for all $\gamma$, it is a saddle point   implying  the emergence of a new (superconducting) global minimum at $T=T_c$. 
 
 Interestingly,  there is a single unstable direction (normal mode) at any $T<T_c$ for $\gamma\gtrsim 2.7$ and for $\gamma\lesssim 0.2$. In other words, there is an orthonormal basis in the configuration space of the free energy, such that  it increases along all axes except one, along which it decreases. Therefore, a single   mode is responsible for the Cooper instability at all $T$
 in the $\gamma$ model with these values of $\gamma$. The same is true in the BCS theory~\cite{barankov0,osc}, which corresponds to $\gamma=0$. It is an  amplitude (Higgs) mode, because  the phase of $\Delta_n$ is zero for it.   
 
 Coherent dynamics of the BCS condensate in response to sudden perturbations can be understood as undamped, underdamped, and overdamped oscillations of the Higgs mode~\cite{osc,barankov,barankov1}. 
 At the first glance, this suggests that  the $\gamma$ model for $\gamma\gtrsim 2.7$ and  $\gamma\lesssim 0.2$  has the same three nonequilibrium phases as the BCS condensate. In Phase I  the order parameter decays to zero at long times, in Phase II it goes to a nonzero   constant, and in Phase III it oscillates persistently and periodically~\cite{swave}. However,  we have to keep in mind that the normal modes we obtain for the $\gamma$ model are the modes of the free energy functional and not of the Hamiltonian dynamics as in the BCS case. Because of this it is far from certain that   the far from equilibrium dynamics of the two systems are indeed similar.

 \subsection{Arbitrary  $\gamma$}
 
 \begin{figure}[!htb]
	\centering
	 \includegraphics[width=1.0\columnwidth]{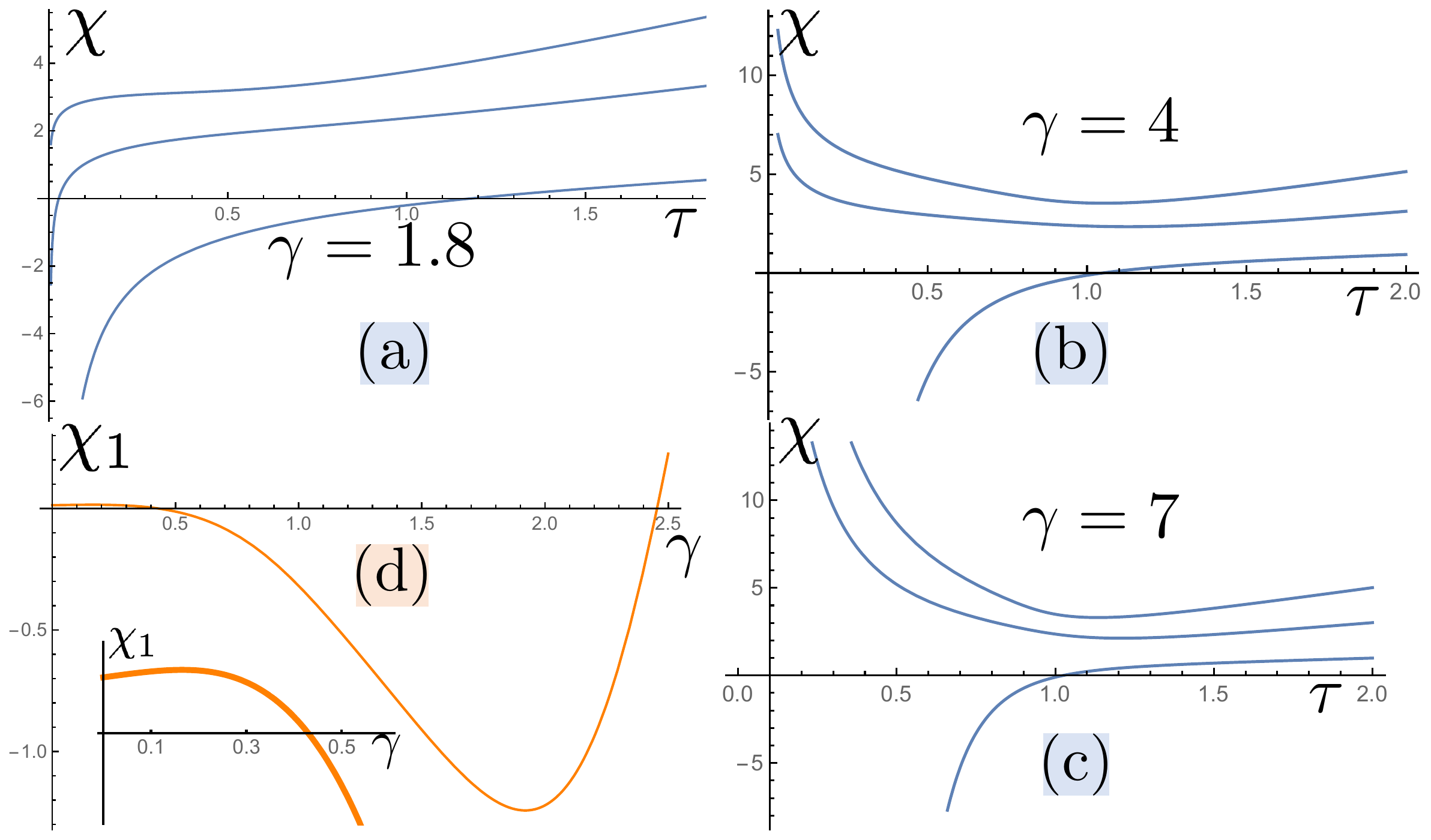}
	\caption{Three lowest eigenvalues   of the second derivative matrix $X$  vs.  reduced temperature $\tau$ for three
	different values of $\gamma$ [(a), (b), (c)] and the second lowest eigenvalue $\chi_1$ 
	as a function of $\gamma$ at $\tau=0.01$ (d). The inset in (d) magnifies the  small $\gamma$ region of the 
	  graph to show that $\chi_1(\gamma)$ becomes positive again at small $\gamma$. Eigenvalues of $X$ determine the stability of the normal state -- it is stable when all of them are positive and unstable otherwise. For any $\gamma$ there is a critical temperature $\tau_c$ below which the normal state is unstable. When $0.2\lesssim \gamma\lesssim 2.7$ two eigenvalues are negative at $\tau\to0$.  Otherwise,  there is a single negative eigenvalue at all $\tau<\tau_c$. }
	\label{eigfig}
\end{figure}

 As usual, to determine the type of the stationary point, we need to   expand the free energy  to quadratic order around it. It is convenient to work with the free energy~\re{freesym} formulated on the positive Matsubara axis. Expanding \eref{freesym} around the normal state, we obtain a quadratic form
\beg
\delta\bar f= \tau\!\!\! \sum_{n,m=0}^\infty \theta_n X_{nm}\theta_m=\tau \bm\theta^T X\bm \theta,
\label{quadratic}
\en
where $\delta\bar f=\bar f - \bar f_\mathrm{n}$, $\bar f_\mathrm{n}$ is the  non-dimensionalized  normal state free energy,   $\bm\theta$ is a column vector with components $\theta_n$, and the second derivative matrix $X$ (the Hessian) reads
\beg
\begin{split}
\tau^{\gamma-1} X_{nn}&=\tau^\gamma\left(n+\frac{1}{2}\right) + \sum_{k=1}^n \frac{1}{k^\gamma}-\frac{1}{2(2n+1)^\gamma},\\
\tau^{\gamma-1} X_{n\ne m}&=-\frac{1}{2(n+m+1)^\gamma}-\frac{1}{2|n-m|^\gamma}.
\end{split}
\label{A}
\en
 The normal state is a minimum when all eigenvalues of $X$ are positive.  It becomes a saddle point   and a transition to the superconducting state takes place when one of the eigenvalues of $X$ vanishes. In a mechanical interpretation, where $\delta\bar f$ is
 the potential energy of a system of 1D point particles with coordinates $\theta_n$ and equal mass, 
 eigenvalues and eigenvectors of $X$ have the meaning of squares of the frequencies and normal modes of small oscillations around the normal state, respectively. The normal state is unstable when one of the frequencies becomes imaginary. 
 
 To analyze the eigenvalues of  $X$ numerically, we truncate it to a finite $L\times L$ matrix. The size $L$ should be such that the Zeeman field overwhelms the interaction near the ends of the chain as it   does at  large Matsubara frequencies in an infinite chain. The first line in \eref{A} shows that this requires $L\gg\tau^{-\gamma}$ for $\gamma>1$ and $(L\tau)^\gamma\gg 1$ for $\gamma<1$. For comparison, the critical dimensionless temperature is $\tau_c\approx 1$ for $\gamma\ge2$ (see below).  It increases with decreasing $\gamma$ and diverges, $\tau_c\sim \gamma^{-\frac{1}{\gamma}}$ as $\gamma\to 0^+$~\cite{2mats}.
 
  Figure~\ref{eigfig} shows three lowest eigenvalues $\chi_0 < \chi_1< \chi_2$ of $X$ as functions of the reduced temperature $\tau$ for  several $\gamma$ and the second  eigenvalue $\chi_2$ as a function of $\gamma$ at a fixed temperature $\tau=0.01<\tau_c$. We observe that: (a)  $\chi_0$ is negative below  a certain value of $\tau$ and positive above it  for all $\gamma$, i.e.,  there is a superconducting transition for any  $\gamma$; (b) $\chi_2$ is positive at all $\gamma$ and temperatures; and (c) $\chi_1$ changes sign twice as a function of $\gamma$ at low temperatures. 
  
  Point (c) comes as a surprise --   the second eigenvalue $\chi_1$ also becomes negative at very low temperatures for $0.2\lesssim \gamma\lesssim 2.7$. Outside of this interval of $\gamma$  only one eigenvalue is negative for all $T<T_c$. It is interesting to see whether  
   this means that the dynamics of  the Cooper instability for $\gamma\gtrsim 2.7$ and  $\gamma\lesssim 0.2$  are as in the BCS model, where there is a single negative eigenvalue as well. Suppose we prepare the system in the normal state and then suddenly turn on the interaction (interaction quench). The BCS
  order parameter first grows exponentially  with the growth exponent  set by the negative eigenvalue~\cite{barankov0,tsyp}. Nonlinear effects stop the growth at some point after which the order parameter oscillates persistently and periodically. The fact that  oscillations occur with a single basic frequency is a consequence of  having only one unstable direction.

\subsection{Large $\gamma$}
\label{largegsubs}

The problem of finding the large $\gamma$ asymptotic behavior of $T_c$ as well as of the eigenvalues and eigenvectors of $X$ is exactly solvable.
First, it is not difficult to see that at $\gamma=\infty$ the critical temperature $\tau_c=1$ ($T_c=\frac{g}{2\pi}$).
Indeed, the  interaction term in   \eref{nonf}   is proportional to $\tau^{2-\gamma}$. At $\gamma=\infty$   the interaction vanishes for $\tau>1$. Then, the system simply minimizes the Zeeman term as shown in Fig.~\ref{spinsfig}(a), i.e., we are in the normal state, where $\theta_n=\frac{\pi}{2}-\frac{\pi}{2}\mbox{sgn}(\omega_n)$.   For $\tau<1$   the interaction diverges.  The sharp domain wall   at the origin
(the jump from $\theta_{-1}=\pi$ to $\theta_0=0$) now costs infinite energy as the interaction is ferromagnetic and favors parallel spin alignment.  It is more advantageous to spread the jump in $\theta_n$ over a large energy interval, i.e., the domain wall softens~\cite{bcinf}. In fact, in this limit all spins at finite $\omega_n$ are along the $x$-axis and $\bar f=0$ for $\tau<1$~\cite{consider}.

Now let us evaluate the leading order asymptotic behavior of $T_c$ and other observables for large $\gamma$.
Near the transition the system is close to the normal state. We  therefore expand the free energy~\re{nonfpos}   around the normal state  to the second order,  
\begin{equation}
\delta\bar f=\tau\sum_{n=0}^\infty \bar\omega_n\theta_n^2
+\frac{\tau^{2-\gamma}}{2}\sum_{n=0}^\infty
(\theta_{n+1}-\theta_n)^2- \tau^{2-\gamma} \theta_0^2.
\label{norm}
\end{equation}
The notations here the same as in \eref{quadratic}.

Finding  eigenvalues of the Hessian $X$ is equivalent to finding the stationary points of $\bar f_\chi=\bar f - \chi \tau \sum_n \theta_n^2$. Setting the derivative of $f_\chi$ with respect to $\theta_n$ to zero, we obtain one equation for $n\ge1$,
\begin{equation}
\theta_{n+1}=2\left[ \tau^\gamma
\left(n+\frac{1}{2}\right)-\tau^{\gamma-1}\chi+1\right]\theta_n-\theta_{n-1},
 \label{rl}
\end{equation}
and another one for $n=0$,
\begin{equation}
\theta_1=[\tau^\gamma-1-2\tau^{\gamma-1}\chi]\theta_0.
 \label{r0}
\end{equation} 
Similar   to a second order linear differential equation, the recurrence relation~\re{rl} has two linearly independent solutions. Equation~\re{r0} provides one boundary condition. The second boundary condition is $\theta_n\to 0$ for $n\to\infty$, see \eref{sym}. 

Observe that \eref{rl}
is the recurrence relation for Bessel functions,
\beg
Z_{\alpha+1}(x)=\frac{2\alpha}{x} Z_\alpha - Z_{\alpha-1},
\en
with $x=\tau^{-\gamma}$ and $\alpha= n+\frac{1}{2}-\tau^{-1}\chi+ \tau^{-\gamma}$. The Bessel function $Z_\alpha(x)$ that goes to zero as $\alpha\to\infty$ is $J_\alpha(x)$ -- the Bessel function of the first kind. Therefore, the normal mode with eigenvalue $\chi$ is
\beg
  \theta_{n\chi} =J_{n+\frac{1}{2}+\tau^{-\gamma}-\chi \tau^{-1}}(\tau^{-\gamma}),
\label{nmode}
\en
up to a normalization  constant. The boundary condition~\re{r0} now determines the eigenvalues $\chi$ at temperature $\tau$,
\begin{equation}
\frac{ J_{\frac{3}{2}+\tau^{-\gamma}-\chi
\tau^{-1}}(\tau^{-\gamma})}{J_{\frac{1}{2}+\tau^{-\gamma}-\chi
\tau^{-1}}(\tau^{-\gamma})}=\tau^{\gamma}-1-2\tau^{\gamma-1}\chi.
\label{gen}
\end{equation}
We use this formula to determine the critical temperature   and derive the  Landau free energy near the critical point.


\section{Critical temperature}
\label{crit_T_sec}

As discussed above, the superconducting transition temperature $\tau_c=\frac{2\pi T_c}{g}$ corresponds to  one of the eigenvalues $\chi$ of the   matrix $X$ crossing zero. Setting $\chi=0$ in \eref{gen}, we obtain an equation for $a=\tau_c^\gamma$,
\begin{equation}
\frac{ J_{\frac{3}{2}+a^{-1}}(a^{-1})}{J_{\frac{1}{2}+a^{-1}}(a^{-1})}=a-1.  
\label{tc}
\end{equation}
Numerically, we find that this equation has a unique solution 
\beg
a\approx1.1843,
\en
 and therefore
\beg
 T_c(\gamma\to\infty)=\left[a+ O(2^{-\gamma})\right]^{\frac{1}{\gamma}} \frac{ g}{2\pi}.
\label{Tc1}
\end{equation}
The correction $O(2^{-\gamma})$ to $a$  is due to interactions with  non-nearest-neighbor   spins   which  we neglected. This formula gives the leading term in the asymptotic expansion of $T_c(\gamma)$  around $\gamma=\infty$. Note that $T_c(\infty)=\frac{g}{2\pi}$ in agreement with our reasoning at the beginning of  Sec.~\ref{largegsubs}. 

\renewcommand{\arraystretch}{1.6}
\setlength{\arrayrulewidth}{0.8pt}
\setlength{\abovecaptionskip}{-4pt}
 
\begin{table}[htb]
\begin{tabularx} {0.43\textwidth} { 
   >{\raggedright\arraybackslash}X 
  >{\raggedright\arraybackslash}X 
   >{\raggedright\arraybackslash}X    >{\raggedright\arraybackslash}X }
\hline 
$\gamma$ & $\frac{T_c(\gamma)}{g}$  & $\frac{T_c(\gamma\to\infty)}{g}$& $\frac{T_c(\gamma)-T_c(\gamma\to\infty)  }{ T_c(\gamma\to\infty)}$\\
\hline 
2 &0.183& 0.173& $6\%$\\
3 & 0.171 & 0.168 & $2\%$\\
4 & 0.1669 & 0.1660 & $0.5\%$\\
8 & 0.16258& 0.16256 & $0.01\%$\\
20 & 0.160506897 &  0.160506899 & $-10^{-6}\%$\\
\hline
\end{tabularx}
\vspace{16pt}
\caption{Values of the superconducting transition temperature $T_c(\gamma)$ for various $\gamma$ versus the large $\gamma$ asymptote~\re{Tc1}. The agreement is reasonable already for $\gamma=2$. The relative error -- the last column of the table -- is roughly
$\frac{2^{-\gamma}}{2\gamma}$ consistent with $O(2^{-\gamma})$ term in \eref{Tc1}.}
\label{tctable}
\end{table}

In Table~\ref{tctable}, we compare the large $\gamma$ asymptote~\re{Tc1}  with numerically exact values of $T_c$ for several $2\le \gamma\le 20$. The agreement is reasonable already at $\gamma=2$. To determine $T_c$ numerically, we diagonalize the Hessian $X$ in \eref{A} as a function of the reduced temperature $\tau=\frac{2\pi T}{g}$ and compute the value of $\tau$ at which the lowest eigenvalue of $X$
crosses zero.

\section{Thermodynamics   near the transition}
\label{thermo_sec}

Near $T_c$ we need to keep only the unstable mode $\theta_{n\chi_0}$ whose eigenvalue $\chi_0$ changes sign at the transition.
 Then, $\theta_n= \epsilon  \theta_{n  \chi_0}$, where the amplitude of the unstable mode $\epsilon$ is our order parameter. The free energy $\bar f$  expanded near the transition to order $\epsilon^4$ is the Landau free energy from which various thermodynamic properties, such as the jump in the specific heat, thermodynamic critical field $H_c$ etc., follow. 

At the transition, the unstable mode [\eref{nmode} with $\chi=0$ and $\tau=\tau_c$] is
\beg
  \theta_{n0}=J_{n+\frac{1}{2}+a^{-1}}(a^{-1}),
\label{nmode1}
\en
where we used $\tau_c^\gamma = a$. This function is positive for all $n\ge 0$. It  decays with $n$ monotonically and very quickly, approximately as $n^{-n}$.  Its values at the first ($n=0$) and second ($n=1$) Matsubara frequencies contribute $96.66\%$ and    $3.28\%$    to $\sum_{n=0}^\infty   \theta_{n0}^2$. The superconductivity is therefore confined to few small Matsubara frequencies, mostly to $\pm \pi T$, consistent with the short ranginess  of the interaction, cf.~\cite{2mats}. In terms of spins, the   domain wall that develops below $T_c$ (see Fig.~\ref{spinsfig}) measures only a couple of sites long.

To obtain the Landau free energy, we expand   \eref{nonf} to quartic order in $\theta_n$ and substitute $\theta_n= \epsilon  \theta_{n  \chi_0}$. The quadratic part simplifies since $\theta_{n  \chi_0}$ is the eigenstate of matrix $X$ with eigenvalue $\chi_0$ and we find
\beg
\delta\bar f=\chi_0 \epsilon^2 \tau_c b_2 + \epsilon^4 \tau_c^2 b_4.
\en
 As $\chi_0\sim\epsilon^2$ at the minimum with respect to $\epsilon$, it is sufficient to evaluate the coefficients at $\chi_0\epsilon^2$ and $\epsilon^4$   at $\tau=\tau_c$ -- corrections to them in $\delta\tau=\tau-\tau_c$ contribute terms of order $\epsilon^6$ to the free energy. 
 Then, $b_2=\sum_n   \theta_{n0}^2\approx 6.074\times 10^{-2}$ and  similarly we determine $b_4\approx 8.445\times 10^{-4}$. 
 
 We also need $\chi_0$  as a function of $\delta\tau$. Expanding \eref{gen} in $\chi$ and $\delta\tau$ around $\chi=0$ and $\tau=\tau_c$, we find $\chi_0= h\gamma\delta\tau$, where $h\approx 0.5339$.  It is convenient to redefine the order parameter as
 $\tau_c\epsilon^2 =\eta^2 |\psi|^2$, where $\eta^2=\frac{h b_2}{2 b_4}$. We have
 \beg
\delta\bar f=R\left [\gamma \delta\tau |\psi|^2+\frac{|\psi|^4}{2}\right],\quad R\approx 0.6226.
\label{landau}
\en
We made $\psi=|\psi| e^{i\phi}$ complex to restore the arbitrary overall phase $e^{i\phi}$ of $\Delta_n$ which we set to one until now. Since $\theta_{n\ge0}$ is small, \esref{spintogap} and \re{spintheta} imply $\Delta_n= \omega_n\theta_n$ for $\omega_n>0$ and therefore
\beg
\begin{split}
\Delta(\omega_n)=\eta|\psi| a^{-\frac{1}{\gamma}}\omega_n J_{n+\frac{1}{2}+a^{-1}}(a^{-1})e^{i\phi},\quad \mbox{$\omega_n>0$},\\
\Delta(-\omega_n)=\Delta(\omega_n),\quad \eta\approx 19.20.
\end{split}
\en
Note that $\Delta(\omega_n) e^{-i\phi}>0$ in agreement with the theorem mentioned earlier that  $\Delta(\omega_n)$ must be  nonnegative     at the minimum of the free energy up to an overall phase.

Let us also restore the units in the Landau free energy~\re{landau},
\beg
f_L\equiv f- f_\mathrm{n}=R \nu_0 g^2 \left [\frac{2\pi\gamma}{g} (T-T_c) |\psi|^2+\frac{|\psi|^4}{2}\right],
\label{conden}
\en
where $f_\mathrm{n}$ is the normal state free energy density.

Minimizing the condensation energy~\re{conden} with respect to $|\psi|^2$, we find
 \begin{subequations}
\begin{eqnarray}
|\psi|=  \left[\frac{2\pi\gamma}{g}\right]^{1/2} (T_c-T)^{1/2},\\
 f-f_\mathrm{n}=-  2\pi^2 R \nu_0   \gamma^2  (T-T_c)^2.\label{con1}
\end{eqnarray}
\end{subequations}
From here the jump in the specific heat $c=-T\frac{\partial^2 f}{\partial T^2}$, i.e., the difference between superconducting  and normal state specific heats at $T=T_c$ is
\beg
\delta c=c_\mathrm{s}-c_\mathrm{n}=2\pi R\nu_0 g \gamma^2 a^{\frac{1}{\gamma}},
\label{Cjump}
\en
where we used \eref{Tc1}. For $\gamma=2$ this formula gives $\delta c \approx 24 \nu_0 g$, which is not too far from the exact answer (rounded to two significant digits)
$\delta c =17 \nu_0 g$~\cite{carbotte} considering that $\gamma=2$ is well outside of  $\gamma>3$ range where our large $\gamma$ theory is supposed to be accurate.

The thermodynamic critical field $H_c$ is the magnetic field above which the energy cost $\frac{H^2}{8\pi}$ of expelling the magnetic field (Meissner effect) exceeds the energy gain~\re{con1} due to superconductivity~\cite{grosso}.  We find
\beg
H_c= 4\pi^{3/2} \gamma \sqrt{R\nu_0} (T_c-T).
\en
Both $|\psi|$ and $H_c$ have the usual mean-field scaling with $(T_c-T)$ for a scalar theory.

\section{Heat capacity above and  below $T_c$}
\label{ns_sec}

Here we evaluate the normal state specific heat  $c_\mathrm{n}$ for $\gamma>2$ and the specific heat $c_\mathrm{s}$ in the superconducting state just below $T_c$. It turns out that $c_\mathrm{n}$ is negative in a   temperature interval $(T_c, T_\mathrm{n})$ above $T_c$ for \textit{any} $\gamma\ge2$.  We will discuss the significance of this later in this section.

We saw above that in the normal state $S_n^z=\mbox{sgn}(\omega_n)$.  The free energy density in terms of the spin chain is $f=\nu_0 T H_s$. The Zeeman term  in \eref{sch2}  contributes
\beg
f^Z_\mathrm{n}=- 2\pi \nu_0T\!\!\! \sum_{n=-\infty}^\infty |\omega_n| =- 8\pi^2 \nu_0T^2 \sum_{n=0}^\infty \left(n+\frac{1}{2}\right)
\label{fZ0}
\en
to the normal state free energy.
  Looking back at the derivation of the spin chain~\cite{spinchain}, we see that up to a $T$-independent  constant, $f^Z_\mathrm{n}$ must be  the $\eps_F\to\infty$ limit of the free energy of the noninteracting Fermi gas~\cite{landau},
 \beg
  f_\mathrm{n}^{(0)}=u_0\eps_F-\frac{1}{3}\pi^2 \nu_0 T^2,
  \label{fn0}
 \en
 where $u_0$ is a constant that depends only on the number of spatial dimensions, e.g., $u_0=3/5$ in 3D. 

Summations in \eref{fZ0} diverge.  The reason is that we took the limit $\eps_F\to\infty$ when deriving the spin chain and the Fermi gas free energy~\re{fn0} diverges in this limit. However,  this affects only the $T$-independent part, which is of no interest to us here. The standard
way to deal with this divergence is to apply the Poisson summation formula to \eref{fZ0}  discarding the $T$-independent part. We obtain
\beg
f^Z_\mathrm{n}=-\frac{1}{3}\pi^2 \nu_0 T^2.
\label{fZ}
\en
It is instructive to also derive this answer using the zeta function regularization technique~\cite{kleinert}.   Recall the definition of the Hurwitz zeta function
\beg
\zeta(s, p)=\sum_{n=0}^\infty \frac{1}{(n+p)^s}.
\en
We interpret  the second summation in \eref{fZ}   as $\zeta(-1,\frac{1}{2})$ and since $\zeta(-1,\frac{1}{2})=\frac{1}{24}$,
we obtain \eref{fZ} with this approach too.

Now note that ${\bm S_n}\cdot {\bm S_m}-1= \mbox{sgn} (\omega_n\omega_m)-1$ vanishes when $\omega_n$ and $\omega_m$ have the same sign and is equal to $-2$ otherwise. This observation allows us to rewrite the interaction part of the free energy as
\beg
\begin{split}
f_\mathrm{int}=&\nu_0 g^\gamma(2\pi T)^{2-\gamma}\sum_{l=1}^\infty \frac{l}{l^\gamma}=\\
&=4\pi^2 \nu_0 \zeta(\gamma-1) \left(\frac{g}{2\pi}\right)^\gamma T^{2-\gamma}.
\end{split}
\label{fint}
\en
Here we reduced the summation over $n$ and $m$ to a single sum over $l=n+m+1$ using $\omega_n-(-\omega_m)\propto (n+m+1)$ and the fact that there are $l$ ways to choose $n$ and $m$ for a given   $l$.

 Adding \eref{fint} to \eref{fZ}, we obtain the normal state free energy up to a $T$-independent constant 
\beg
f_\mathrm{n}=-\frac{1}{3}\pi^2 \nu_0 T^2+ 4\pi^2 \nu_0 \zeta(\gamma-1) \left(\frac{g}{2\pi}\right)^\gamma T^{2-\gamma}
\en
and
\beg
c_\mathrm{n}(T)=\frac{2\pi^2\nu_0}{3} T\left[ 1-\left(\frac{T_\mathrm{n}}{T}\right)^{\!\!\gamma} \right],
\label{nstsh}
\en
where
\beg
T_\mathrm{n}=[3(\gamma-1)(\gamma-2)\zeta(\gamma-1)]^{\frac{1}{\gamma}} \frac{g}{2\pi}> T_c.
\en
The inequality $T_\mathrm{n}(\gamma)>T_c(\gamma)$ holds for all $\gamma>2$ because the difference $T_\mathrm{n}(\gamma)-T_c(\gamma\to\infty)$ grows with $\gamma$ and exceeds $0.25 T_c$ already at $\gamma\to2^+$~\cite{limit}, while $T_c(\gamma\to\infty)$  given by \eref{Tc1} underestimates $T_c$ by  6\% or less as seen from Table~\ref{tctable}.

 \begin{figure}[!tb]
	\centering
	 \includegraphics[width=1.0\columnwidth]{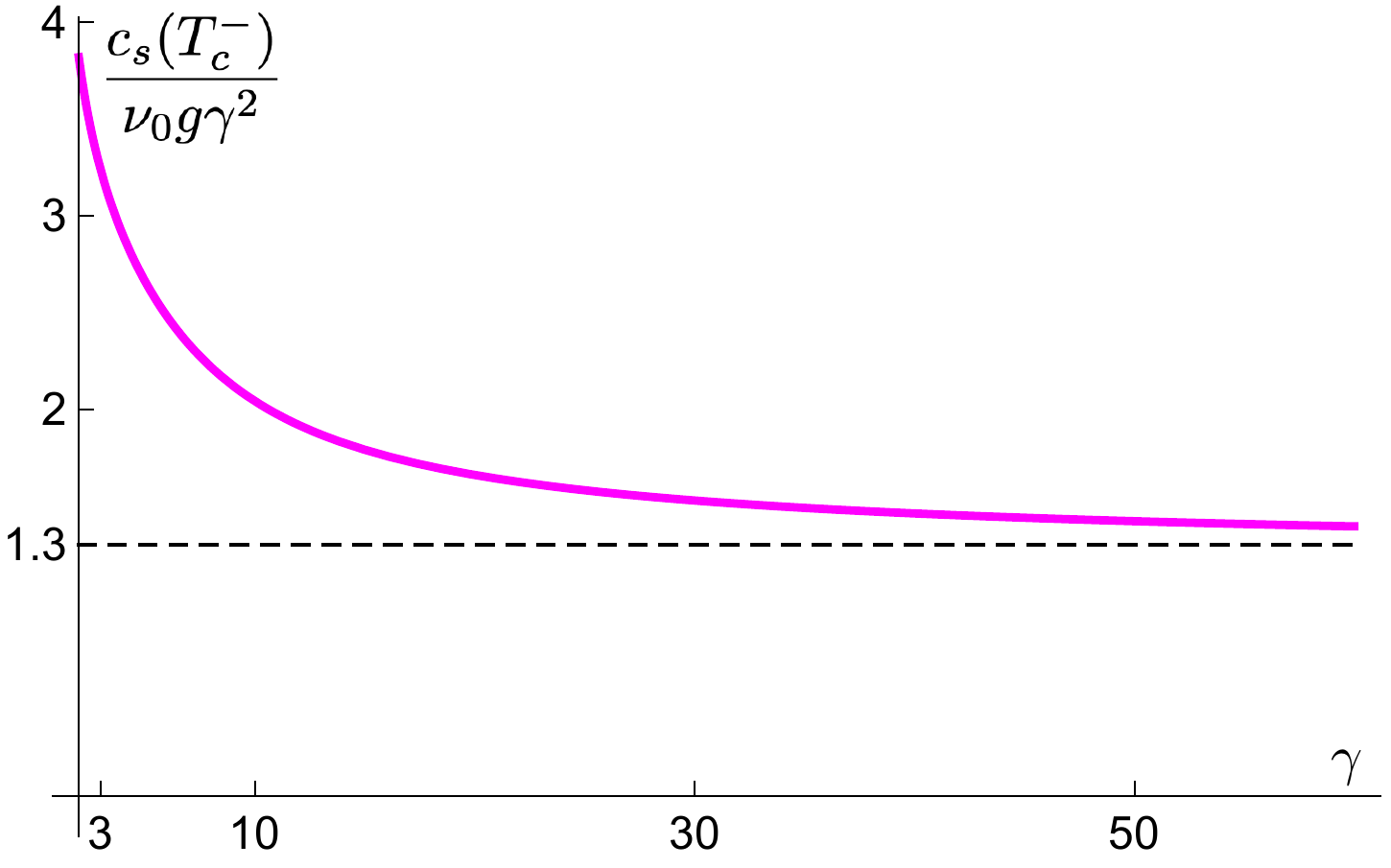}
	\caption{Specific heat $c_\mathrm{s}$ in the superconducting state just below $T_c$ as a function of $\gamma$,  normalized by $\nu_0 g\gamma^2$. 
	We see that $c_\mathrm{s}$ is positive at $T=T_c^-$ for all $\gamma>3$.  Since $c_\mathrm{s}(T)$ is continuous,  it  will remain positive for a finite temperature range below $T_c$. Note also that $c_\mathrm{s}(T_c^-)\approx 1.3 \nu_0 g \gamma^2$ at large $\gamma$.  }
	\label{spheatTcfig}
\end{figure}

Therefore, the heat capacity is negative for $T_c<T<T_\mathrm{n}$. Note that $T_\mathrm{n}\to T_c$ as $\gamma\to\infty$. Nevertheless, there is a sliver of temperature  where the heat capacity is negative. The interpretation of  this depends on the origin of the $\gamma$ model.
Any subsystem that does not interact with other subsystems and has negative heat capacity is thermodynamically unstable~\cite{landau}.  
One scenario is that the quasiparticles are ill defined, they do not form a   Fermi liquid, and  therefore  the stationary point we started with (the solution of the gap equation) is not the global minimum of the total free energy.  Another scenario is that the effective fermion-fermion interaction $V(\omega_l)$ changes with temperature, so that $V(\omega_l)=\frac{g^\gamma}{|\omega_l|^\gamma}$ below $T_c$ and  something else above $T_c$. In other words, the $\gamma$ model kicks in only below $T_c$. This scenario is in principle possible when the bosons that mediate the interaction are  collective excitations of the fermions themselves as is the case for many $\gamma <1$. Then, the superconducting transition     modifies the bosonic propagator and  therefore $V(\omega_l)$. However, if the interaction is mediated by phonons or other true bosons, the $\gamma$ model is unphysical for $\gamma\ge 2$ at least in a certain range of temperatures, see also Ref.~\cite{breakEli} where  we addressed this issue for phonon mediated electron-electron interactions ($\gamma=2$).

 Due to the jump at $T_c$, the specific heat  becomes positive in the superconducting state for a range of temperatures  $T_-< T \le T_c$. We encountered this situation before in the $\gamma=2$ case, where the superconducting state was free of the pathologies of the normal state -- the opening of the gap stabilized the system. To evaluate  the specific  heat $c_\mathrm{s}$   just below $T_c$, i.e., at $T=T_c^-$, we use \esref{Cjump} and \re{nstsh} and
 \beg
 c_\mathrm{s}(T_c^-)=c_\mathrm{n}(T_c^-)+\delta c.
 \en
 We plot  $c_\mathrm{s}(T_c^-)$ normalized by $\nu_0 g\gamma^2$ as a function of $\gamma$ 
 in Fig.~\ref{spheatTcfig}. We see that it is positive at $T=T_c^-$ for all $\gamma>3$. By continuity it must also remain positive in a certain finite temperature range $(T_-, T_c^-]$. Note also the large $\gamma$ asymptote $c_\mathrm{s}(T_c^-)\approx 1.3 \nu_0 g \gamma^2.$ We will see in Sec.~\ref{cs_sec} that $T_-> 0$, i.e.,  $c_\mathrm{s}$ becomes negative at low temperatures. This is unlike the $\gamma=2$ case where the specific heat is always positive in the superconducting state and vanishes when $T\to0$~\cite{breakEli} as it should. 
  
 \section{Low temperature properties -- universal gap function and specific heat}
 \label{lowTsec}

We saw that the $\gamma$ model is a superconductor  in thermal equilibrium at $T<T_c$ -- the anomalous averages are nonzero.
 Having addressed its properties     near $T_c$ and  in the normal state, we now turn to the superconducting state at $T\ll T_c$.  
 
 For all $\gamma>3$, the problem of determining the gap function $\Delta(\omega)$ and thermodynamic properties in the low temperature regime reduces to a single parameterless second order ODE up to corrections of relative order $\left(T/g\right)^{r\gamma/3},$ where $r>0$ is  a function of $\gamma$. In this section, we first solve this ODE to evaluate the gap function on the Matsubara axis, the free energy and the specific heat, and then discuss the corrections. Similar to the normal state, the absolute specific heat is negative, but the difference between   the superconducting and normal state specific heats is positive.

 \subsection{Gap function}
 
We   showed in Sec.~\ref{anyg3} that at low temperatures and $\gamma>3$ the gap equation reduces to \eref{ode}.  
Rescaling  the variable $\bar\omega$ as in \eref{resc}, we obtain a \textit{universal} low-temperature gap equation
\beg
 \frac{d^2\theta}{d x^2}= x \sin\theta(x),
 \label{ugap}
 \en
 where
\beg
x= \frac{\omega}{\omega_*},\quad \omega_* =g \frac{[\zeta(\gamma-2)]^{\frac{1}{3}}}{2^{\frac{1}{3}} } \left(\frac{g}{2\pi T}\right)^{\!\!\frac{\gamma}{3}-1}\!\!\!\!\!\!\!\!\!\!. 
\label{chi}
\en
The energy constant $\omega_*$ is the coupling $g$ rescaled by a factor that depends on $g/T$ and $\gamma$. \eref{ugap} is universal in that it is parameterless and therefore  independent of $\gamma$, $g,$ and $T$ -- all
dependence on these parameters is in the energy scale $\omega_*$. 

 \begin{figure}[!htb]
	\centering
	 \includegraphics[width=1.0\columnwidth]{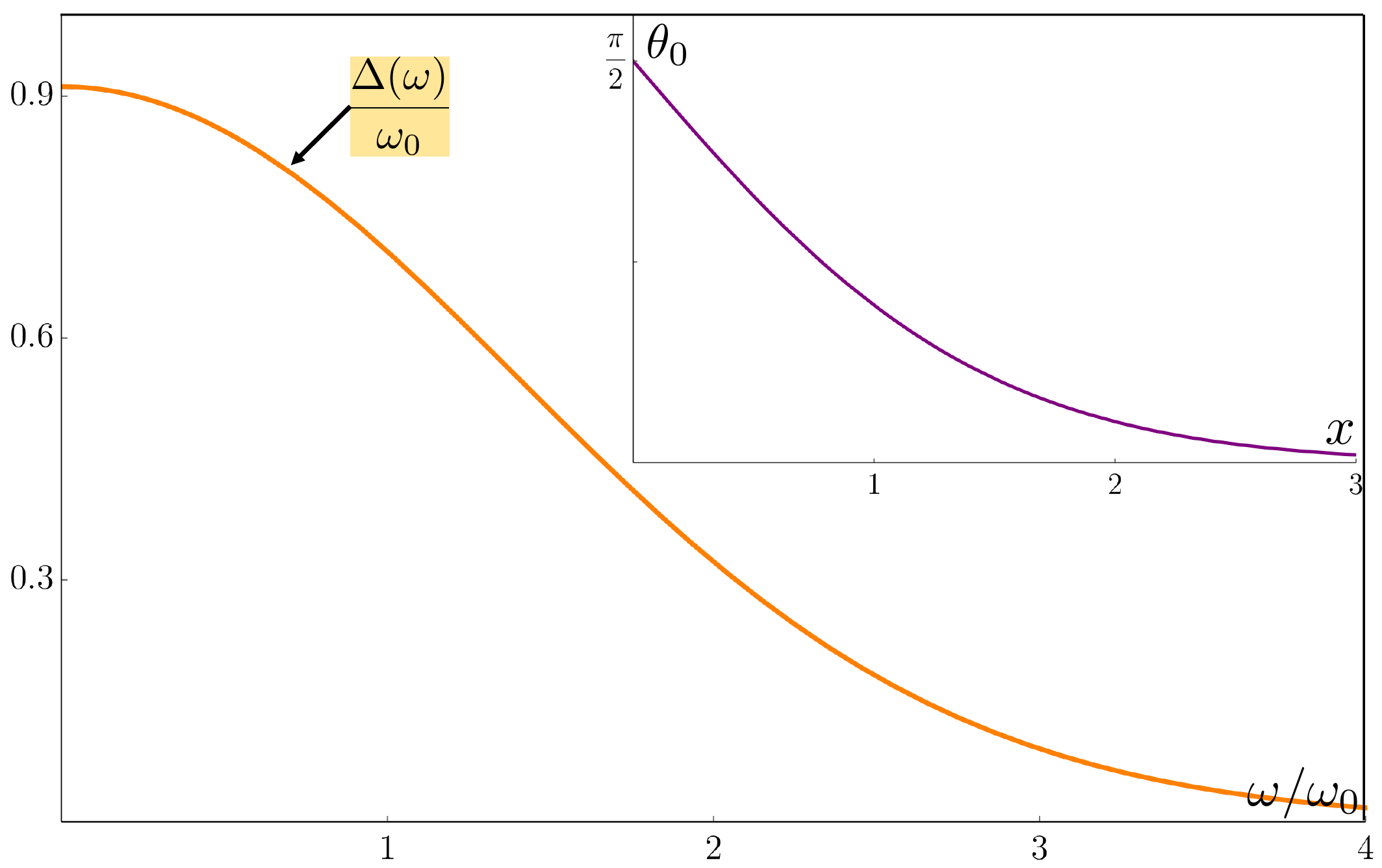}
	\caption{Universal gap function $\Delta(\omega)$    and the corresponding angle $\theta_0(x)$ spins make with the $z$ axis (inset). Gap functions $\Delta(\omega_n)$    for all $\gamma>3$ and all values of the coupling  $g$ collapse onto this plot in the limit $T\to0$ as long as both $\Delta$ and $\omega$ are measured in units of the energy constant $\omega_*$ given by \eref{chi}. }
	\label{deltafig}
\end{figure}

Recall from Sec.~\ref{anyg3} that $\theta(0)=\frac{\pi}{2}$. Further, the requirement $\Delta_n\ge0$ discussed in the beginning of Sec.~\ref{stpt} together with \esref{spintheta} and \re{deltatheta} imply $0\le \theta_n\le\frac{\pi}{2}$  for  $\omega_n>0.$
 Therefore, for $x\ge0$ we have 
\beg
\theta(0)=\frac{\pi}{2},\quad \theta(+\infty)=0,\quad 0\le\theta(x)\le\frac{\pi}{2}.
\label{bc}
\en
 In the Appendix, we show that under the conditions~\re{bc} there exists a unique solution
$\theta_0(x)$ of the nonlinear differential equation~\re{ugap} and note that $\theta_0(x)$ is  real analytic  on the entire $x$-axis by standard theorems of the theory of ODEs. This is a parameterless function which we plot  in the inset to Fig.~\ref{deltafig} for $x\ge0$.  Observe also that since $\theta(x)\to\pi-\theta(-x)$ leaves \eref{ugap} invariant and by uniqueness, $\theta_0(x)$ must map into itself under this transformation, i.e.,  $\theta_0(x)=\pi-\theta_0(-x)$.

Having determined $\theta_0(x)$, we know  the gap function  for all low temperatures, $g$, and $\gamma>3$. 
Indeed, \eref{deltatheta} implies
\beg 
 \frac{ \Delta(\omega)}{\omega_*}=Y\!\left(\frac{\omega}{\omega_*}\right), \quad Y(x)= x \tan \theta_0(x).
\en
[Recall that we extended $\Delta(\omega_n)$ to $\Delta(\omega)$ defined on the entire real $\omega$ axis in Sec.~\ref{anyg3}.]
This equation provides the leading small $T$ asymptotic behavior of $\Delta(\omega)$ for any $g$ and $\gamma$.
Graphically,  plots of the gap function $\Delta(\omega_n)$  vs.  Matsubara frequency $\omega_n$ for any $\gamma$ and $g$  tend to the same universal curve shown in Fig.~\ref{deltafig} as $T\to0$, when both $\Delta(\omega_n)$ and $\omega_n$ are measured in units of $\omega_*$.

It is straightforward to work out the expansion of $\Delta(\omega)$ at small $\omega$ and its   large $\omega$ asymptote. Both depend on a single constant that needs to be determined numerically. At large $\omega/\omega_*$, the angle $\theta_0$ is small
and $Y(x)\approx x\theta_0(x)$.
Equation~\re{ugap} becomes the Airy equation. Its solution that goes to zero at infinity is the Airy function of the first kind, $\theta_0(x)\propto\mathrm{Ai}(x)$. Therefore at large $\omega/\omega_*\equiv x$
\beg
\frac{ \Delta(\omega)}{\omega_*}\approx 4.58 \mathrm{Ai}(x)\approx 1.29 x^\frac{3}{4} \exp\biggl(- \frac{   2x^\frac{3}{2} }{3}\biggr),
\en
where  we determined the constant of proportionality 4.58 numerically and used the known asymptotic expansion for $\mathrm{Ai}(x)$~\cite{stegun}. 

Similarly, at small $x=\omega/\omega_*$
\beg 
 \frac{ \Delta(\omega)}{\omega_*}\approx 0.91-0.10 x^2+0.05 x^4+ O(x^6).
 \en
 In particular, recalling the definition~\re{chi} of $\omega_*$, we find
 \beg
 \Delta(0)= 0.72 g  [\zeta(\gamma-2)]^{\frac{1}{3}}  \left(\frac{g}{2\pi T}\right)^{\!\!\frac{\gamma}{3}-1}.
 \en
 Since this result is for $\gamma>3$, $\Delta(0)$ diverges as $T^{1-\frac{\gamma}{3}}$ for ${T\to0}$~\cite{andrey_ratio}. For fixed $T$
 and $\gamma\to 3^+$ we have
 \beg
 \frac{2\Delta(0)}{T_c}\approx 8.47 |\gamma-3|^{-\frac{1}{3}},
 \en
where we took the value of $T_c$ for $\gamma=3$ from Table~\ref{tctable}. Reference~\cite{andrey_ratio} found the same answer, but for $\gamma\to 3^-$ and with a prefactor $4\pi$ instead of 8.47. 
Note that unlike in the BCS theory, $2\Delta(0)$ here is not the gap in the spectrum~\cite{notgap}.

\subsection{Low $T$ free energy and specific heat}
\label{cs_sec}

To evaluate the free energy $\bar f$, we use \eref{freelow}. As discussed in Sec.~\ref{ns_sec}, the free energy contains a diverging $T$-independent constant because we took the limit $\eps_F\to\infty$. This divergence is present for superconducting states as well, since in these states $S_n^z\to\mathrm{sgn}(\omega_n)$ when $|\omega_n|\to\infty$, same as in the normal state.
To isolate it, we subtract and add the noninteracting part of the normal state free energy    $f_\mathrm{n}^Z$   given by \eref{fZ0}, i.e., $ f= f -  f_\mathrm{n}^Z+ f_\mathrm{n}^Z$.  The difference $ f -  f_\mathrm{n}^Z$ is finite and we have already evaluated the  $T$ dependence of $f_\mathrm{n}^Z$  in Sec.~\ref{ns_sec}.

Taking the difference  $\bar f - \bar f_\mathrm{n}^Z$, where $\bar f_\mathrm{n}^Z= f_\mathrm{n}^Z/(\nu_0 g^2)$, corresponds to the replacement $S^z\to S^z-\mbox{sgn}(x)$ in the Zeeman term  in \eref{freelow}, since $S^z(x)=\mbox{sgn}(x)$ in the normal state. Using also $S^z=\cos\theta$, $d\bm S/dx=d\theta/dx\equiv\theta'$, and the symmetry property~\re{sym} of $\theta(x)$, we obtain the  non-dimensionalized free energy in the superconducting state at $T\ll T_c$ as
\beg
\begin{split}
\bar f_\mathrm{s}= &\frac{[\zeta(\gamma-2)\tau^{3-\gamma}]^{\frac{2}{3}}}{2^{\frac{2}{3}}} \!\!\! \int\limits_0^\infty \!\! dx \! \left\{ 2x (1-\cos \theta_0) +(\theta'_0)^2\right\}\\
&-\frac{\tau^2}{12},\quad\mbox{where}\quad \tau=\frac{2\pi T}{g},\quad \bar f =\frac{f}{\nu_0 g^2}.
\end{split}
\en
 Substituting the numerical solution for $\theta_0(x)$ discussed in the previous subsection into the integral, we find
 \beg
 \bar f_\mathrm{s}=1.11 [\zeta(\gamma-2)]^{\frac{2}{3}}\tau^{2-\frac{2\gamma}{3}}-\frac{\tau^2}{12}.
 \en
 The specific heat  $c_\mathrm{s}=-T\frac{\partial^2 f}{\partial T^2}$ in the superconducting state is
 \beg
 c_\mathrm{s}=\frac{2\pi^2\nu_0}{3}T \left[1-\left(\frac{T_s}{T}\right)^{\!\!\frac{2\gamma}{3}}\right],
 \en
 where
 \beg
 T_s=\left[1.48(\gamma-3)(2\gamma-3)\right]^\frac{3}{2\gamma} \left[\zeta(\gamma-2)\right]^\frac{1}{\gamma}\frac{g}{2\pi}.
 \en
 It is evident that $c_\mathrm{s}$  becomes negative as $T\to 0$ for any $\gamma>3$. 
 
This indicates  that the $\gamma$ model is pathological for $T\to0$ just as it is above $T_c$, see the discussion at the end of Sec.~\ref{ns_sec}. Again, there cannot be an unambiguous explanation of this pathology without the knowledge of the system Hamiltonian.  To fix it,  we need to modify the effective fermionic action~\re{postulate}, i.e., the $\gamma$ model itself. One scenario is as follows. Consider the difference between $c_\mathrm{s}$ and
 the normal state specific heat~\re{nstsh} $c_\mathrm{n}$,
 \beg
 c_\mathrm{s}-c_\mathrm{n}=\frac{2\pi^2\nu_0}{3}T \left[\left(\frac{T_\mathrm{n}}{T}\right)^{\!\!\gamma}-\left(\frac{T_s}{T}\right)^{\!\!\frac{2\gamma}{3}}\right].
 \label{difflowT}
 \en
This is positive for $T<T_\mathrm{n}^3/T_s^2$. A straightforward numerical analysis shows that $T_\mathrm{n}^3/T_s^2>T_c$. Therefore, this difference is always positive  when $T<T_c$.  Suppose the modification of the fermionic action is such that it introduces a new order parameter independent of  $\bm S_n$. Further, suppose  this adds terms to $H_s$  that depend on the new order parameter and not on   $\bm S_n$ [see
\esref{sch} and~\re{freespin}]. Then, it does not affect the superconducting transition and the gap equation, but adds 
$T$-dependent terms to the free energy and to the specific heat. The change in the specific heat must be the same for $c_\mathrm{s}$ and $c_\mathrm{n}$. Provided the new $c_\mathrm{n}$ is positive, so is the new $c_\mathrm{s}$, because \eref{difflowT} remains valid. This scenario is favorable for the results of this section in the sense that it fixes the pathology while leaving  them intact with the exception of the answers for $\bar f_\mathrm{s}$ and $c_\mathrm{s}$.  However, note that  $\bar f_\mathrm{s}-\bar f_\mathrm{n}$ and $c_\mathrm{s}-c_\mathrm{n}$ do not change.

\subsection{Accuracy of the local approximation}

In this section, we solved the gap equation on the Matsubara axis and evaluated the specific heat at low temperatures for $\gamma>3$, by replacing the gap equation~\re{eligapeq}, which we copy here for convenience, 
\beg
 \omega_n\sin\theta_n= g^\gamma \pi T \sum_{m\ne n} \frac{\sin(\theta_m-\theta_n)}{|\omega_m-\omega_n|^\gamma},
 \label{eligapeq1}
 \en
  with the differential equation~\re{ugap}.  Now let us investigate the accuracy of this local approximation.
  
 To derive \eref{ugap}, we expanded $\sin(\theta_m-\theta_n)$ in $(\omega_m-\omega_n)$ to second order, see \eref{repl}. The contribution from odd powers of $(\omega_m-\omega_n)$ to the right hand side of \eref{eligapeq1} cancels after summation over $m$.
 Therefore, the error comes from terms of order 4 and higher.  The Lagrange error bound $M(\omega_m-\omega_n)^4/4!$ 
then gives an upper bound on the difference between $\sin(\theta_m-\theta_n)$ and our  approximation to it. Here $M$ is the maximum value of $d^4\theta_n/d\omega_n^4$. 
 
 Let us split the summation in \eref{eligapeq1} into two parts: $|m-n|< L$ and $|m-n|\ge L$, where $L$ is to be determined. For $|m-n|\le L$ we use the Taylor series expansion of $\sin(\theta_m-\theta_n)$ to third order plus the   Lagrange error bound. For $|m-n|> L$, we replace $\sin(\theta_m-\theta_n)\to 1$, which provides an upper bound for the error from neglecting these terms. Pulling  out an overall factor of $2(2\pi T)^{-\gamma}$, we obtain
 \beg
 \begin{split}
 \frac{(2\pi T)^2}{2} \frac{d^2\theta_n}{d\omega_n^2}\zeta(\gamma-2) -\frac{(2\pi T)^2}{2} \frac{d^2\theta_n}{d\omega_n^2}  \sum_{k=L}^\infty k^{2-\gamma}\\
 + \frac{(2\pi T)^4}{4!} M\sum_{k=1}^L k^{4-\gamma}+\sum_{k=L}^\infty k^{-\gamma}.
 \end{split}
 \label{error}
 \en
 The first term leads to \eref{ugap} after the change of variables~\re{chi}. The remaining three terms are the error. To estimate the  second derivative of $\theta(\omega_n)$ and $M$,   we use the numerical solution of  \eref{ugap} plotted in Fig.~\ref{deltafig}. For example, $d^2\theta/d\omega^2=\omega_*^{-2} d^2\theta/dx^2$ with $0\le d^2\theta/dx^2<0.6$ and $M\approx 1.4 \omega_*^{-4}$. We replace the summations over $k$ in \eref{error}  with integrals and then minimize the error with respect to $L$. As $L$ turns out to be large, corrections due to the replacement of the summations with integrals are negligible. In this way, we obtain that the relative error [magnitude of the ratio  of the sum of the last three terms in \eref{error} to the first term] is
 \beg
 \mathrm{RE}\sim \left(\frac{2\pi T}{g}\right)^{\!\!\frac{r\gamma}{3}}\!\!\!\!\!\!,\quad
 r=\left\{
 \begin{array}{ll}
 \gamma-3, & \mbox{for $3<\gamma\le 5,$}\\
 2, & \mbox{ for $\gamma>5$}.\\
 \end{array}\right.
 \en

\section{Conclusions}
\label{conclusion_sec}

In this paper, we studied the thermodynamics of a system of fermions near a quantum critical point with extremely retarded interactions of the form   $V(\omega_l)=(g/|\omega_l|)^\gamma$, where $\omega_l$ is a bosonic Matsubara frequency. The case $\gamma=2$ of this $\gamma$ model corresponds to the strong coupling limit of the Eliashberg theory, which is intermediately retarded.  Extreme retardation at generic $T$ kicks in for $\gamma\gg 1$ and for $\gamma>3$ at $T\to0$.  Note that for  $\gamma> 2$, the $\gamma$ model  is a model without  a Hamiltonian and is defined through a nonlocal effective Euclidian action only.

The $\gamma$ model shows two phases:  normal  and superconducting.  Similarly to the Eliashberg theory, the order parameter is the frequency dependent gap function $\Delta(\omega_n)$, where $\omega_n$ is the fermionic Matsubara frequency. We determined the superconducting transition temperature $T_c$ and the order parameter $\Delta(\omega_n)$ near $T_c$. The amplitude $\psi$ of $\Delta(\omega_n)$ serves as   the Landau order parameter.   We expanded the free energy functional to order $|\psi|^4$ near the transition to obtain the Landau free energy, from which we  derived the jump in the specific heat and the thermodynamic critical field.  These answers are asymptotically exact in the limit $\gamma\to\infty$. We also evaluated the normal state specific heat $c_\mathrm{n}$ for arbitrary $T$ and $\gamma>2$.

Next, we turned our attention to the properties of the $\gamma$ model at low temperatures. We proved that the global minimum of the free energy is unique (nondegenerate). We derived the universal gap equation, which is a parameterless second order
ODE, and determined the scaling of $\Delta(\omega)$ with $T$, $g$, and $\gamma$ for all $g$, $\gamma>3$, and $T\to0$. Building on this, we obtained explicit expressions for the   free energy and specific heat in the superconducting state for this range of parameters.
We also evaluated $2\Delta(0)/T_c$ and found that it is finite for $T>0$ and $\gamma>3$, but diverges as $|\gamma-3|^{-1/3}$ for 
$\gamma\to 3^+$ and as $T^{1-\gamma/3}$ for $T\to0$.
These results are exact for any $\gamma>3$ at $T\to0$. 

Note that ``exact'' and ``asymptotically exact'' here and elsewhere in this paper mean exact for 
the $\gamma$ model defined by the effective action~\re{postulate} in the thermodynamic and $\eps_F\to\infty$ limits. In this regime, fluctuational corrections to the spin chain are negligible and it is at zero effective temperature, i.e., its ground state [which is determined by the gap equation~\re{eligapeq}]    captures all thermodynamical properties that we evaluated.

We found that the $\gamma$ model is thermodynamically unstable for $\gamma\ge2$. Its specific heat is negative above $T_c$ for $\gamma\ge 2$ and also at $T\to0$ for $\gamma>3$. We saw  in  an earlier paper~\cite{breakEli}  that  when this model is understood as an effective description of   the phonon mediated electron-electron interaction ($\gamma=2$), this instability implies the emergence of a new order   above $T_c$. 
The new phase breaks the lattice translational symmetry and invalidates the $\gamma$ model at least in a certain temperature range.
For other $\gamma$, a microscopic Hamiltonian is similarly necessary to resolve this issue. An interesting open problem  is therefore to construct  classes of physical many-body Hamiltonians that correspond to the $\gamma$ model with arbitrary  $\gamma>2$.

\begin{acknowledgments}

We thank Ar. Abanov, A. V. Chubukov, G. Kotliar, and T.-H. Lee for helpful  discussions.

\end{acknowledgments}

\onecolumngrid 

\medskip

\appendix*

 \section{Existence and uniqueness of solutions to the universal gap equation}
\label{ODE}

In this appendix we show that under the conditions~\re{bc} the universal gap equation~\re{ugap} has a unique solution.
Uniqueness is easier to show than existence, so we start with uniqueness, then discuss existence of a solution.
We also offer a section about the properties of solutions of \eref{ugap} when conditions \re{bc} are dropped.

\subsection{Uniqueness of a solution satisfying conditions~\re{bc}}

Suppose that both $\theta_0(x)$ and $\tilde\theta_0(x)$ are twice continuously differentiable solutions of the ODE \re{ugap} 
that satisfy the conditions~\re{bc}.
 Then $\Theta(x):= \theta_0(x) - \tilde\theta_0(x)$ vanishes when $x\to 0$ from the right, and when $x\to\infty$. 
 Moreover, $\Theta(x)$ is also twice continuously differentiable, and its second derivative is given by
\begin{equation}
\frac{d^2}{dx^2}\Theta(x) = x \left(\sin\big[\theta_0(x)\big]-\sin\big[\tilde\theta_0(x)\big]\right),\quad x\in (0,\infty).
\label{ThetaEQ}
\end{equation}
We multiply \eref{ThetaEQ} by $\Theta(x)$, 
integrate from $x=0$ to $x=\infty$,  integrate by parts on the left-hand side, and obtain
\begin{equation}
0\geq - \int_0^\infty |\Theta^\prime(x)|^2 dx 
= 
\int_0^\infty x |\Theta(x)|^2\, \frac{\sin\big[\theta_0(x)\big]-\sin\big[\tilde\theta_0(x)\big]}{\theta_0(x)-\tilde\theta_0(x)} dx 
\geq 0,
\label{sandwichINEQ}
\end{equation}
where the prime denotes derivative with respect to $x$.
 The second inequality in \re{sandwichINEQ} follows from the fact that $\sin\theta$ is an increasing 
function for $\theta\in[-\pi/2,\pi/2]$. 
 Since both inequalities in \re{sandwichINEQ} are strict if $\theta_0(x_0)\neq \tilde\theta_0(x_0)$ for 
some $x_0 >0$ (and therefore in some open neighborhood of $x_0$), the
only option compatible with \re{sandwichINEQ} is: $\theta_0(x)=\tilde\theta_0(x)$ for all $x\geq 0$.

Thus uniqueness holds.

\subsection{Existence of a solution satisfying conditions~\re{bc}}

The uniqueness proof only shows that there cannot exist more than one twice continuously differentiable solution 
	to \eref{ugap} that satisfies the conditions \re{bc}, while leaving open whether a solution exists at all. 
		 In this subsection we show that such a solution does exist, indeed.

 We now consider the initial value problem for \eref{ugap}, with initial data 
\begin{equation}
\theta(0) = \frac{\pi}{2}; \qquad \theta^\prime(0) = \alpha,
\label{iDATA}
\end{equation}
and we treat $\alpha$ as a real parameter that we exhibit explicitly in the solution to the initial value
problem, written as $\theta(x;\alpha)$. 
 Our goal is to show that there exists at least one particular value $\alpha_0$ (possibly not unique, in this section)
such that the pertinent solution $\theta(x;{\alpha_0})$ of the initial value problem satisfies the 
remaining conditions in
\re{bc}; i.e., $\theta(x;{\alpha_0})$ vanishes as $x\to\infty$, and $\theta(x;{\alpha_0})$ 
takes values only in $[0,\pi/2]$ for $x\geq 0$. 
 In this case we may identify $\theta(x;{\alpha_0})$ with a sought-after solution $\theta_0(x)$ of \eref{ugap} under
the conditions \re{bc}.

 We begin by noting that the Picard-Lindel\"of   theorem~\cite{Kamke} guarantees that
for each $\alpha$ the initial value problem for \eref{ugap} with initial data \re{iDATA}
has a unique twice continuously differentiable solution $\theta(x;\alpha)$, which exists for all $x\in\mathbb{R}$.
 The regularity follows from the facts that the function $x\sin \theta$ is 
continuous in both $x$ and $\theta$. The uniqueness and global character of such a solution 
follow from the additional feature that the derivative of $\sin \theta$ is uniformly bounded in absolute value.
 In fact, $\theta(x;\alpha)$ is analytic in $x$; this follows from the fact that the function $x\sin \theta$ is 
analytic in both $x$ and $\theta$; see Ref.~\cite{davis}.
 Thus it remains to show that there is a solution $\theta_0(x)$ that converges to 0 as $x\to\infty$, and that this solution
does not take values outside of the interval $[0,\pi/2]$ when $x\geq 0$.

 Since we demand that the solution $\theta_0(x)\leq \pi/2$, with $\theta_0(0)=\pi/2$, it now follows from \eref{ugap}
that \emph{a necessary condition for the existence of such a solution is that the initial slope} $\alpha<0$. 
 Indeed, for $\theta \in (0,\pi)$ and $x>0$, the right-hand side of \eref{ugap} is $>0$, so that the 
unique solution $\theta(x;\alpha)$ of
\eref{ugap} that satisfies the initial data~\re{iDATA} is convex as long as $\theta(x;\alpha)\in(0,\pi)$. 
 But this means that if there is any $x_0\geq 0$ for which
  $\theta(x_0;\alpha)\in (0,\pi)$ and $\theta^\prime(x_0;\alpha)=0$, 
then $x_0$ is a local minimum point of $\theta(x;\alpha)$, and this
solution will inevitably increase to values $>\pi/2$. 

 On the other hand, since we also demand that the solution $\theta_0(x)\geq 0$, 
its negative initial slope $\alpha$ cannot be too large in magnitude, for it is straightforward to show that
there is some $\alpha_0<0$ such that for $\alpha <\alpha_0$ there is an $x_0>0$ such that
the pertinent solution $\theta(x;\alpha)$ satisfies $\theta(x_0;\alpha)=0$ and $\theta^\prime(x_0;\alpha)<0$. 
(We have recycled the symbol $x_0$ with a new meaning.)
   In this case it follows right away that $\theta(x_0+\epsilon;\alpha) < 0$ for some $\epsilon>0$.
 To see that there is such an $\alpha_0<0$, recall that $\sin\theta \leq 1$, so that we obtain the estimate
\beg
\theta^{\prime\prime}(x;\alpha)\leq x,
\en
and integrating this estimate twice for the stipulated initial data we find that
\beg
\theta(x;\alpha) \leq \tfrac{\pi}{2} + \alpha x + \tfrac16 x^3.
\en
The cubic polynomial at the right-hand side may have no, or one, or two positive roots, depending on $\alpha$. 
 If there is at least one positive root, let $x_*$ denote either the unique positive
root or the smaller of the two positive roots.  
 Such a root $x_*$ exists if and only if $\alpha +\frac12 x_*^2 \leq 0$.
 Setting $\alpha_* = - \frac12 x_*^2$, our cubic problem becomes
$\tfrac{\pi}{2} -\frac12 x_*^3 + \frac16 x_*^3 =0$, viz. $x_* = \left(\frac{3\pi}{2}\right)^{\frac13},$
which returns
\beg
 \alpha_* = -\tfrac12 \left(\tfrac{3\pi}{2}\right)^{\frac23}.
\en
 And so, when $\alpha \leq \alpha_*$ the cubic upper bound to the solution $\theta_\alpha(x)$ of 
our initial value problem, i.e., \eref{ugap} with initial conditions~\re{iDATA}, vanishes at $x_*$ with a slope $\leq 0$.
Hence, $\theta(x;\alpha)$ itself must have a zero at some $x_0<x_*$ when $\alpha\leq \alpha_*$. 
 Moreover, $\theta^\prime(x_0;\alpha) <0$.
 For suppose $\theta^\prime(x_0;\alpha)=0$;
then both $\theta(x;\alpha)$ and its first $x$ derivative would vanish at $x_0$, and by the uniqueness of the solution
of the second-order initial value problem formulated with these data at $x_0$, the function $\theta(x;\alpha)$
would have to vanish identically, which is a contradiction to the fact that $\theta(0;\alpha) = \frac{\pi}{2}$.
 So $\theta^\prime(x_0;\alpha)<0$, and since $\theta(x_0;\alpha)=0$, it follows that $\theta(x+\epsilon;\alpha)<0$,
which violates the required lower bound 0 for $\theta_0$.
 It follows that there is some $\alpha_0$ with $\alpha_* < \alpha_0 < 0$ such that 
\emph{a further necessary condition for the existence of the desired solution} $\theta_0(x)$
\emph{is that the initial slope} $\alpha \geq \alpha_0$.

Now consider what happens to $\theta(x;\alpha)$ if we start with
$\alpha = \alpha_*$ and continuously increase $\alpha$ from there.
  As just discussed, the solution $\theta(x;{\alpha_*})$ to the initial value problem \re{ugap}, \re{iDATA}
has a smallest positive zero at $x_0(\alpha_*) <x_*$, and $\theta^\prime\big(x_0(\alpha_*);\alpha_*\big)<0$.  
 Moreover, $\theta^\prime(x;\alpha_*)<0$ for all $x\in[0,x_0(\alpha_*)].$ 
Indeed, if there was some $x_\star<x_0$ with
$\theta^\prime(x_\star;\alpha_*)=0$, then $\theta(x_\star;\alpha_*) \in (0,\pi/2)$, and as discussed above,
$x_\star$ would be a local minimum point of $\theta(x;{\alpha_*})$. This solution would 
increase for $x>x_\star$ to values $>\pi/2$ -- in contradiction to the cubic upper bound we derived. 
 Thus $\theta(x;{\alpha_*})$ decreases monotonically from the value $\frac{\pi}{2}$ at $x=0$ to the
value 0 at $x=x_0(\alpha_*)$.   
 Now increase $\alpha$ continuously above $\alpha_*$.
 It is easy to see [just formally integrate \eref{ugap} twice, using \eref{iDATA}] 
that for all $0<x\leq x_0(\alpha)$ one has $\frac{\partial}{\partial\alpha}\theta^\prime(x;\alpha)>0$
and $\frac{\partial}{\partial\alpha}\theta(x;\alpha)>0$.
 Thus the zero $x_0(\alpha)$ moves continuously to the right as $\alpha$ increases.
 Moreover, as long as $x_0(\alpha)<\infty$, the function $\theta(x;\alpha)$ reaches its first 
zero at $x_0(\alpha)$ with a nonzero negative slope, $\theta^\prime\big(x_0(\alpha);\alpha\big)<0$.
 This follows from the already made observation that 
$\theta(x;\alpha)$  must be identically zero if it vanishes at a finite location $x_0(\alpha)$
with vanishing slope, in contradiction to the initial data $\theta(0;\alpha)=\frac{\pi}{2}$.
 Thus we can increase $\alpha$ until a value $\alpha_0$ is reached at which $x_0(\alpha_0)=\infty$,
with $\lim_{\alpha\to\alpha_0}\theta^\prime\big(x_0(\alpha);\alpha\big)=0$ (limit from the left). 
 The solution $\theta(x;{\alpha_0})$ is a sought-after solution $\theta_0(x)$.

 This demonstrates the existence of a solution to \eref{ugap} that satisfies \re{bc}.

\subsection{The types of solutions for general initial data}

We briefly discuss the general initial value problem for \eref{ugap} with initial data at $x=0$. 
By the $2\pi$-periodicity of the sine function, it suffices to restrict the discussion to data
\beg
\theta(0)= \vartheta \in [-\pi,\pi];\qquad \mbox{and}\qquad  \theta^\prime(0) = \alpha\in \mathbb{R}.
\en
 For any pair of such initial data $(\vartheta,\alpha)$ there exists a unique 
analytical solution $\theta(x;{\vartheta,\alpha})$ of \eref{ugap} for all $x \geq 0$; cf.~\cite{davis}.
 The purpose of this subsection is to present a mostly 
qualitative and partly quantitative overview of the behavior of these solutions.
 
 To get a more intuitive grasp of the possible solutions $\theta(x;{\vartheta,\alpha})$ 
 it is helpful to note that for $x>0$ the variable transformation
\begin{equation}
\theta(x) = \phi(t),\qquad \mbox{with}\qquad 
t = {\textstyle{2\over 3}} x^{3/2},
\end{equation}
changes \eref{ugap} into
\begin{equation}
\ddot\phi(t) + {\textstyle{1\over 3t}}\dot\phi(t) = \sin\phi(t)
\label{pendulum}
\end{equation}
for $t>0.$ Here we have introduced Newton's dot notation to denote derivatives with respect to $t$.
 Equation~\re{pendulum} describes a damped rigid pendulum, or, equivalently,  
a point mass moving on an upright circle subject to uniform gravity and Newtonian friction, with a friction coefficient
inversely proportional to time $t$~\cite{noworries}.
 We count the angle $\phi$ from the ``up'' position, i.e., $\phi=0$ corresponds to the unstable inverted pendulum equilibrium
 and $\phi=\pi$ to the stable equilibrium. 
Of course, \textsl{a priori} any integer multiple of $2\pi$ may be added to either $0$ or $\pi$ to obtain yet another 
unstable, respectively stable equilibrium solution for \eref{pendulum}, because \eref{pendulum} in itself 
does not restrict $\phi(t)$ to lie in any particular interval such as $[0,\pi/2]$. 

 The damped pendulum  \eref{pendulum} intuitively suggests that
the damping will force the evolution of $\phi(t)$ to converge at $t\to\infty$ to
one of the infinitely many copies of the two possible equilibrium states.
 Most initial data would lead to a (copy of the) stable equilibrium state $\phi_{s}=\pi$.
 However, for each $\vartheta$ there should be a discrete set of slopes $\alpha$ which launch a solution that
asymptotically approaches (a copy of) the unstable equilibrium state $\phi_u=0$.
 The approach to any stable equilibrium is damped oscillatory, while the approach to an unstable
equilibrium is monotonic.
 Copies of the stable pendulum equilibrium have Newtonian energy $=-1$ and the
copies of the unstable one $=+1$.

 There is one caveat to what we just wrote.
 Since the friction coefficient vanishes as $t\to\infty$, it is in principle conceivable that
there also exist solutions that asymptotically approach a dynamical solution of the undamped pendulum equation,
instead of converging to an equilibrium solution.
 We can rigorously rule out an asymptotic approach to a so-called librating solution of the
undamped pendulum equation (see below), but we have not rigorously ruled out an asymptotic
approach to an oscillating undamped pendulum solution; we expect that it does not occur, though.

 In addition to these intuitive insights into the possible types of solution, we offer some exact results.
 
First, one can  show the convergence of
the Newtonian energy of the pendulum evolution
\beg
 E[\phi,\dot\phi](t) := {\textstyle{1\over 2}} \big|\dot\phi(t)\big|^2 + \cos\phi(t).
\en
  The energy is monotonically decreasing with   $t$ due to friction.
 Indeed,  
\beg
{\textstyle{d\over dt}} E[\phi,\dot\phi](t) = \dot\phi(t)\big[\ddot\phi(t) - \sin\phi(t)\big] = - {\textstyle{1\over 3t}}|\dot\phi(t)|^2 \leq 0,
\label{Edot}
\en
and for any $0<t<\infty$ the right-hand side in \eref{Edot} is ``$<0$'' if $\dot\phi\neq 0$.
 Since $E$ is bounded below, it  must converge as $t\to\infty$.
 However,  it may or may not converge to its minimum value $E_{\min} = -1$.

 We remark that convergence of $E[\phi,\dot\phi](t)$ to a constant  as $t\to\infty$
does not in itself imply convergence of $\phi(t)$ as $t\to\infty$.
After all, in the absence of friction the energy is conserved, hence trivially  converges to a constant.
At the same time, $\phi(t)$ keeps oscillating or librating forever in this case, unless $\phi(t)$ is one
of the two equilibrium states.
 With the help of our energy dissipation identity \re{Edot}
we can rule out an asymptotic approach to a librating solution, though.
 Namely, since $-1\leq \cos \phi \leq 1$, we can extract from \eref{Edot} the two-sided bounds
\beg\label{sandwich}
-1 + C_1 t^{-2/3} \leq E[\phi,\dot\phi](t) \leq 1 + C_2 t^{-2/3},
\en
with $C_1$ and $C_2$ some positive constants.
 Since any asymptotically librating solution has energy $E \geq 1+\epsilon$ for some $\epsilon >0$, it follows
from the upper bound in \eref{sandwich} that the only possibilities are the asymptotic approach to the stable
equilibrium ($E=-1$), to the unstable one ($E=1)$, or to an undamped oscillating pendulum solution ($-1<E<1$).

 Second, whenever $\phi(t)$ does converge to one of the equilibrium points of the damped rigid pendulum,
i.e., to a zero of $\sin\phi$, linearization about that equilibrium yields an accurate approximation at large $t$.
 Letting $\xi(t)$ denote the deviation of $\phi(t)$ from the equilibrium point, and dropping 
terms nonlinear in $\xi$, we find that $\xi(t)$ is (asymptotically) a solution of 
one of the following two ODEs, viz.,
\beg
\ddot\xi(t) + \tfrac{1}{3t}\dot\xi(t) = \pm \xi(t).
\label{AiryEQ}
\en
 These are Bessel differential equations. Their solutions for positive $t$ map into the Airy function $\mathrm{Ai}$
in the original variables.
 More precisely, for large enough $x$ the function $\theta(x;{\vartheta,\alpha})$
approaches the asymptotic form $C_1 + C_2\;$Ai$(\pm x +C_3)$.
 The constant $C_1$ is an integer multiple of $\pi$, and the ``$+$'' sign is to be chosen when an unstable
pendulum equilibrium is approached monotonically, while the ``$-$'' sign pertains to the oscillatory 
approach to a stable pendulum equilibrium.
 The constants $C_1$, $C_2$, and $C_3$ depend on the initial data.


\begin{thebibliography}{99}
\bibitem{alternative}  This derivation of the spin chain     holds everywhere in the configuration space (not only at the stationary points)  when $\Omega\to0$. A different derivation is available for general $\Omega$. See Secs.~II and VII in Ref.~\cite{spinchain}.
\bibitem{grand}  To be more precise,  taken at its minimum $f$ is the  grand potential per volume as we work at fixed chemical potential. Nevertheless, we colloquially refer to it as the free energy or  free energy density.
\bibitem{spinflip} Here we do not consider the spin-flip saddle points~\cite{spinchain} where the sign of the right hand sides of \esref{spintogap} flips for certain $n$.
\bibitem{confused} Not to be confused with the real frequency axis for the original complex frequency $z$. In terms of $z$, Matsubara frequencies $\omega_n$ correspond to discrete points $z_n=i\omega_n$ on the imaginary axis and  the ``real $\omega$ axis'' -- to imaginary $z$ axis, $z=i\omega$.
\bibitem{bcinf} At any finite $\gamma$,  spins approach their normal state positions when $\omega_n\to\pm\infty$, because $\Delta_n\to0$ in these limits, see \eref{spintogap}. For $\gamma=\infty$ we impose this as boundary conditions at infinity.
\bibitem{consider} Consider $\theta_n$ that changes slowly with $n$. The interaction part of the free energy~\re{nonf} is then proportional to $\int_{-\infty}^\infty dn (\theta')^2$, where $\theta'=d\theta/dn$. The Zeeman term is negligible at any finite $n$. The interaction is to be minimized subject
to the constraint $\int_{-\infty}^\infty dn \theta'=-\pi$. Take $\theta'=-\pi/L$ for $|n|<L/2$ and $\theta'=0$ otherwise. This satisfies the 
constraint and it is also clear that the interaction attains its absolute minimum when $L\to\infty$.
\bibitem{notgap} The spectral gap is twice the solution of the equation $\Delta(\omega)=\omega$ on the real axis.
\bibitem{noworries} There is no complication at $t=0$ due to this $t$-dependent damping term because we already know that the
initial value problem is well-defined in the $x$ variable  at $x=0$. It is therefore sufficient to consider  the   problem in the $t$ variable with data at some $t_0>0.$
\end{thebibliography}
\end{document}